\pdfoutput=1
\documentclass[conference]{IEEEtran}
\IEEEoverridecommandlockouts

\usepackage[T1]{fontenc}
\usepackage{cite}
\usepackage{amsmath,amssymb}
\usepackage{graphicx}
\usepackage{booktabs}
\usepackage{tabularx}
\usepackage{multirow}
\usepackage{xcolor}
\usepackage{tikz}
\usetikzlibrary{arrows.meta,positioning,fit,calc}
\usepackage{pgfplots}
\usepgfplotslibrary{statistics}
\pgfplotsset{compat=1.17}
\usepackage{placeins}
\usepackage{pifont}
\usetikzlibrary{shapes.geometric}
\tikzset{
  fb/.style={draw=black!70, rounded corners=1.5pt, align=center, font=\scriptsize, inner sep=2.2pt,
             minimum height=0.5cm, text width=#1, execute at begin node={\hyphenpenalty=10000\exhyphenpenalty=10000}},
  fb/.default=7.4cm,
  cN/.style={fill=gray!9}, cB/.style={fill=blue!10}, cG/.style={fill=green!13}, cR/.style={fill=red!13},
  cY/.style={fill=yellow!25}, cP/.style={fill=violet!13}, cW/.style={fill=white},
  ar/.style={-{Stealth[length=3pt]}, draw=black!70, thin},
  lab/.style={font=\tiny\itshape, inner sep=1pt},
  ttl/.style={font=\scriptsize\bfseries, inner sep=1pt},
  grpbox/.style={draw=black!70, rounded corners=2pt, inner sep=3pt},
  db/.style={cylinder, shape border rotate=90, aspect=0.22, draw=black!70, align=center, font=\scriptsize,
             minimum height=0.85cm, text width=1.5cm}
}
\usepackage{listings}
\usepackage[hidelinks]{hyperref}
\hypersetup{pdftitle={A Consolidated Deployment Edition of a Microservice Core-Banking Platform: Composing Services by Reference into Three Hosts}, pdfauthor={Tomisin Fagbola}, pdfkeywords={modular monolith, service composition, core banking, differential testing, fault injection, AWS Fargate}}

\newcommand{\code}[1]{\texttt{\small #1}}
\newcolumntype{L}{>{\raggedright\arraybackslash}X}
\newcommand{\yes}{\ding{51}}
\newcommand{\no}{--}
\newcommand{\term}[1]{\emph{#1}}

\newcommand{\RpsreadThirtyTwoOld}{167}
\newcommand{\RpsreadThirtyTwoNew}{379}
\newcommand{\RatioreadThirtyTwo}{2.3}
\newcommand{\RatioCIreadThirtyTwo}{2.1--2.6}
\newcommand{\PninetyfivereadThirtyTwoOld}{789}
\newcommand{\PninetyfivereadThirtyTwoNew}{135}

\newcommand{\RpswriteEightOld}{13.3}
\newcommand{\RpswriteEightNew}{19.5}
\newcommand{\RatiowriteEight}{1.5}
\newcommand{\RatioCIwriteEight}{1.3--2.0}

\newcommand{\RpswriteThirtyTwoOld}{12.4}
\newcommand{\RpswriteThirtyTwoNew}{24.2}
\newcommand{\RatiowriteThirtyTwo}{2.0}
\newcommand{\RatioCIwriteThirtyTwo}{1.6--2.5}

\newcommand{\NonOkwriteThirtyTwoOld}{39}
\newcommand{\NonOkwriteThirtyTwoNew}{19}
\newcommand{\PerfRounds}{10}
\newcommand{\MemPeakOld}{2,071}
\newcommand{\MemPeakNew}{995}
\newcommand{\OverTasks}{accounts 524 MiB in 6 of 10 rounds}
\newcommand{\StartOld}{3.2--8.5}
\newcommand{\StartNew}{1.2--6.4}
\newcommand{\ConsIdleNew}{39}
\newcommand{\ConsIdleOld}{24}
\newcommand{\PerfRows}{%
Reads, 8 clients & 169 [157, 173] & 437 [398, 516] & 2.58 [2.30, 3.22] \\
\quad p95, ms & 192 [190, 194] & 60.5 [55.7, 65.1] & 0.32 [0.29, 0.34] \\
Reads, 32 clients & 167 [153, 168] & 379 [358, 406] & 2.27 [2.12, 2.59] \\
\quad p95, ms & 789 [754, 844] & 135 [120, 163] & 0.17 [0.15, 0.21] \\
Reads, 64 clients & 154 [133, 170] & 384 [352, 400] & 2.49 [2.11, 2.96] \\
\quad p95, ms & 1,749 [1,525, 1,977] & 279 [274, 286] & 0.16 [0.14, 0.19] \\
Reads, 32, ALB direct & 182 [154, 197] & 414 [391, 467] & 2.27 [1.99, 3.01] \\
\quad p95, ms & 695 [691, 864] & 116 [110, 136] & 0.17 [0.13, 0.20] \\
Postings, 8 clients & 13.3 [11.6, 13.8] & 19.5 [17.3, 24.8] & 1.46 [1.26, 1.98] \\
\quad p95, ms & 1,710 [1,610, 1,786] & 1,361 [1,066, 1,482] & 0.80 [0.60, 0.91] \\
Postings, 32 clients & 12.4 [10.3, 14.2] & 24.2 [23.0, 26.8] & 1.96 [1.65, 2.49] \\
\quad p95, ms & 4,718 [4,386, 10,797] & 3,565 [3,533, 3,609] & 0.76 [0.32, 0.83] \\}
\newcommand{\ResRows}{%
\multicolumn{6}{@{}l}{\emph{Nine-service edition}}\\
accounts & 0.25 / 512 & 107 & \textbf{516 / 524} & 0 & 23 \\
transactions & 0.25 / 512 & 210 & 321 / 470 & 13 & 9 \\
customers & 0.25 / 512 & 83 & 254 / 271 & 0 & 8 \\
users & 0.25 / 512 & 101 & 249 / 284 & 0 & 13 \\
channels & 0.25 / 512 & 98 & 104 / 110 & 0 & 0 \\
background & 0.25 / 512 & 188 & 188 / 203 & 10 & 0 \\
transaction consumer & 0.25 / 512 & 227 & 227 / 237 & 24 & 6 \\
EOD consumer & 0.25 / 512 & 70 & 72 / 72 & 8 & 6 \\
users consumer & 0.25 / 512 & 103 & 103 / 115 & 9 & 9 \\
\emph{Sum of nine, peak} & 2.25 / 4,608 & & 2,071 & & \\
\midrule
\multicolumn{6}{@{}l}{\emph{Consolidated deployment edition}}\\
API host & 1 / 2,048 & 270 & 530 / 548 & 9 & 96 \\
Consumer host & 0.5 / 1,024 & 245 & 276 / 286 & 39 & 15 \\
Background host & 0.5 / 1,024 & 194 & 194 / 207 & 10 & 0 \\
\emph{Sum of three, peak} & 2.0 / 4,096 & & 995 & & \\
\midrule
Gateway (edition runs) & 0.5 / 1,024 & 110 & 114 / 119 & 4 & 20 \\}
\newcommand{\BoxPlots}{%
\addplot+[boxplot prepared={draw position=1, lower whisker=136.4, lower quartile=157.0, median=169.4, upper quartile=173.1, upper whisker=191.3}, fill=gray!45, draw=black, solid] coordinates {};
\addplot+[boxplot prepared={draw position=2, lower whisker=360.0, lower quartile=398.2, median=437.5, upper quartile=515.8, upper whisker=523.0}, fill=blue!35, draw=black, solid] coordinates {};
\addplot+[boxplot prepared={draw position=4, lower whisker=132.6, lower quartile=153.5, median=166.6, upper quartile=168.2, upper whisker=230.4}, fill=gray!45, draw=black, solid] coordinates {};
\addplot+[boxplot prepared={draw position=5, lower whisker=278.6, lower quartile=357.9, median=378.7, upper quartile=405.7, upper whisker=436.3}, fill=blue!35, draw=black, solid] coordinates {};
\addplot+[boxplot prepared={draw position=7, lower whisker=106.7, lower quartile=133.0, median=153.8, upper quartile=170.3, upper whisker=180.6}, fill=gray!45, draw=black, solid] coordinates {};
\addplot+[boxplot prepared={draw position=8, lower whisker=288.7, lower quartile=352.1, median=383.5, upper quartile=399.7, upper whisker=418.6}, fill=blue!35, draw=black, solid] coordinates {};
}

\newcommand{\ConclPerf}{served \RatioreadThirtyTwo{} times the read throughput at lower tail latency, posted 1.5
to 2 times as fast, and used less than half the memory, for an estimated 11 to 24\% saving in service-task compute at list prices (the shared gateway, load balancer
and data stores excluded)}
\newcommand{\PerfSection}{%
\textbf{Method (E3).} The comparison ran in \PerfRounds{} independent rounds per edition. Each round deployed the
edition afresh from the seed snapshot, built the same data with the workload, sampled resources at idle for 15\,s,
warmed every task with 60\,s of reads and 20\,s of postings that were not measured, and then ran six scenarios once
each: a 15\,s pause to let the previous scenario drain, a 10\,s warm-up and a 30\,s
measurement with \code{oha}, while every task's CPU and resident memory were sampled once a second. The four read
scenarios ran first and the two write scenarios after them, each block in a random order seeded by the round number
(pilot rounds showed that a read scenario run straight after a write scenario inherits its backlog, and that
the first scenario of an unwarmed round measures a cold deployment). Within a round
the two editions ran one after the other, and the edition that went first alternated between rounds, so slow drift of
the host falls on both. Reads cycle through 12 GET endpoints across all five modules; writes are balanced two-leg
postings rotating over five accounts. All traffic goes through the gateway except one read scenario sent straight to
the ALB. Table~\ref{tab:perf} gives the median and interquartile range over rounds, and the ratio of the medians with
a 95\% percentile-bootstrap confidence interval \cite{efron1979bootstrap} (10,000 resamples, each edition's rounds
resampled independently). The intervals rest on two assumptions worth stating: that rounds are independent, which
fresh deployments and alternating order support but one shared host does not guarantee; and that ten rounds
describe the distribution well enough, which makes the widest intervals, such as the p95 posting latency at 32 clients
(ratio 0.76, interval 0.32--0.83), the least reliable. We do not report significance tests; the intervals are meant to show how much the ratios moved
between rounds. The tables are generated from the raw per-run results by one script, and Tables~\ref{tab:roundsrps}
and~\ref{tab:roundsp95} in the appendix give every round, so the medians, ranges and ratios can be recomputed. The absolute throughput and latency figures
are measurements of the emulation, not of production AWS; the ratios between the editions are the result.

\begin{table}[t]
\centering\scriptsize
\caption{Throughput (req/s) and p95 latency (ms) under the Fargate task sizes: median [interquartile range] over
\PerfRounds{} rounds, and the ratio edition / nine-service of the medians with its 95\% bootstrap CI.}
\label{tab:perf}
\setlength{\tabcolsep}{1.6pt}
\begin{tabular}{@{}lccc@{}}
\toprule
Scenario & Nine-service & Edition & Ratio [95\% CI]\\
\midrule
\PerfRows
\bottomrule
\end{tabular}
\end{table}

\begin{figure}[t]
\centering
\begin{tikzpicture}
\begin{axis}[width=0.98\columnwidth, height=4.6cm, ymin=0, xmin=0, xmax=9,
  xtick={1.5,4.5,7.5}, xticklabels={8 clients, 32 clients, 64 clients}, ylabel={read req/s},
  tick label style={font=\scriptsize}, label style={font=\scriptsize}, boxplot/draw direction=y]
\BoxPlots
\end{axis}
\end{tikzpicture}
\caption{Read throughput per round; grey: nine-service edition, blue: composed edition (box: interquartile
range and median; whiskers: minimum and maximum over \PerfRounds{} rounds).}
\label{fig:perf}
\end{figure}

\textbf{Reads.} At 32 clients the edition served \RpsreadThirtyTwoNew{} read requests a second against
\RpsreadThirtyTwoOld{} (ratio \RatioreadThirtyTwo, 95\% CI \RatioCIreadThirtyTwo), with a p95 latency of
\PninetyfivereadThirtyTwoNew{}\,ms against \PninetyfivereadThirtyTwoOld{}\,ms. The intervals at 8 and 64 clients and
for the ALB-direct runs exclude a ratio of one in the same direction (Table~\ref{tab:perf}). The cause is visible in
the resource samples (Table~\ref{tab:resources}): five of the twelve read endpoints belong to the accounts module,
and in the nine-service edition the accounts service runs at its 0.25-vCPU quota while the other services have
capacity to spare, so its queue grows and the latency tail lengthens. In the edition the five modules share one vCPU,
so a busy module uses what the idle ones leave. Sending the reads straight to the ALB changes little in either
edition, so the 0.5-vCPU gateway is not the bottleneck at these rates.

\textbf{Writes.} The edition also posted faster: \RpswriteEightNew{} against \RpswriteEightOld{} postings a second
at 8 clients (ratio \RatiowriteEight, 95\% CI \RatioCIwriteEight) and \RpswriteThirtyTwoNew{} against
\RpswriteThirtyTwoOld{} at 32 (ratio \RatiowriteThirtyTwo, CI \RatioCIwriteThirtyTwo), and at 32 clients it rejected
fewer postings with ``could not acquire lock'' (\NonOkwriteThirtyTwoNew\% against \NonOkwriteThirtyTwoOld\%).
Postings are serialised per account by the Redis lock, so the time a posting holds the lock bounds the rate. In the
nine-service edition the transactions service ran at its 0.25-vCPU quota throughout the write scenarios (median
22--24\% of a core), so each posting held the lock longer; the edition's API host used 33--46\% of its one vCPU.
Rates of 13 to 24 postings a second on five accounts are low in absolute terms, a property of the posting path that
both editions share.

\begin{table}[t]
\centering\footnotesize
\caption{Resident memory (MiB) and CPU (\% of one core) per task: idle median; peak as median / maximum over
rounds; CPU at idle and during 32-client reads (medians). Bold: peak above the task's memory.}
\label{tab:resources}
\setlength{\tabcolsep}{2.6pt}
\begin{tabular}{@{}lrrrrr@{}}
\toprule
Task & vCPU / MiB & Idle & Peak & CPU idle & CPU reads\\
\midrule
\ResRows
\bottomrule
\end{tabular}
\end{table}

\textbf{Memory and sizing.} The three hosts of the edition peaked at \MemPeakNew\,MiB together against 4\,GiB
allocated, and the nine services at \MemPeakOld\,MiB against 4.5\,GiB, because each process carries its own runtime,
shared library and caches. In the nine-service edition, tasks whose resident memory exceeded their 512\,MiB size were:
\OverTasks. Resident memory only approximates what a task's memory limit counts, but on Fargate a task that exceeds its
hard limit is stopped, so the accounts service is sized too small for the load measured here. Both consumer deployments use
CPU while idle: the nine-service transaction consumer \ConsIdleOld\% of a core against its 25\% quota, and the
edition's consumer host \ConsIdleNew\% against 50\%, which points to queue polling without a wait.

\textbf{Start-up.} Under their quotas the nine services became healthy \StartOld\,s after start and the edition's
hosts \StartNew\,s after start, across all rounds.

\textbf{Cost (E4).} Table~\ref{tab:cost} estimates the monthly Fargate compute of each edition at eu-west-1 list
prices for Linux/x86 (\$0.04048 per vCPU-hour and \$0.004445 per GB-hour \cite{awsfargatepricing}), with one task of
each kind running for a full month (730\,h). Only the service tasks are priced. The gateway, the ALB, the data stores,
data transfer, logs and storage are the same in both editions or were not modelled, and each task's ephemeral storage
stays within the free 20\,GB. The figures are therefore estimated service-task compute savings, not savings in the total
cost of operating the platform. At the sizes deployed today, the edition costs \$72.08 a month against \$81.09 for the
nine-service edition, 11\% less. At the sizes the measurements call for, the gap widens: the nine-service accounts
service needs 1\,GiB because it exceeded its 512\,MiB, while the edition's background host fits the smallest task
size, so the edition costs \$63.07 against \$82.72, 24\% less. For one copy of the platform this is \$9 to \$20 a
month, and the saving repeats for every environment or tenant that runs a full copy. On Graviton (ARM) list prices
both editions would cost about a fifth less (\$57.67 against \$64.88 at the deployed sizes); we priced this option but
did not measure performance on Graviton.

\begin{table}[t]
\centering\footnotesize
\caption{Estimated monthly Fargate compute, eu-west-1 list prices, 730\,h, one task each. Gateway, ALB, data stores,
transfer, logs and storage excluded.}
\label{tab:cost}
\begin{tabularx}{\columnwidth}{@{}Lrrr@{}}
\toprule
& Nine-service & Edition & Change\\
\midrule
As allocated: vCPU / GiB & 2.25 / 4.5 & 2.0 / 4 & \\
\quad \$ per month & 81.09 & 72.08 & $-11\%$\\
Measured sizing: vCPU / GiB & 2.25 / 5.0$^a$ & 1.75 / 3.5$^b$ & \\
\quad \$ per month & 82.72 & 63.07 & $-24\%$\\
\bottomrule
\end{tabularx}
\\[2pt]{\scriptsize $^a$ accounts raised to 1\,GiB (above 512\,MiB in 6 of 10 rounds). $^b$ background host lowered to 0.25 vCPU / 0.5\,GiB.}
\end{table}

\textbf{Answer to RQ3 (performance and cost).} In the emulation the composed edition served more than twice the
read throughput at a much lower tail latency, posted faster because the posting path was no longer limited to a quarter
of a vCPU, and used less than half the memory. Its estimated service-task compute saving is 11 to 24\% at list prices, with the gateway, the ALB, the data stores,
transfer and logs excluded. These results come
from the emulation and from list prices; throughput and latency on AWS itself were not measured.%
}

\newcommand{\FailReps}{5}
\newcommand{\FailRows}{%
Replacement start to healthy & 5.5\,s [5.0, 6.7] & 4.3\,s [1.2, 5.8]\\
Postings unavailable & 57.9\,s [55.6, 66.4] & 17.2\,s [16.2, 22.1]\\
Reads down: transactions & 40.8\,s [38.3, 46.0] & 17.0\,s [16.1, 21.8]\\
Reads down: accounts, users & 0.0\,s [0.0, 0.0] & 17.2\,s [16.3, 22.0]\\
Reads down: customers$^c$ & 16.2\,s [14.0, 18.3] & 17.2\,s [16.3, 21.9]\\
Acked postings lost (total) & 0 of 184 & 0 of 1083\\
Partial or unbalanced (total) & 0 & 0\\
Committed, not acked (per run) & 0--1 & 0--1\\
Ledger findings (total) & 0 & 0\\}
\newcommand{\OutageRows}{%
Postings hung to client timeout (120\,s) & 1--1 & 1--1\\
Failed with 500 after about 30\,s & 1--1 & 1--1\\
Rejected, ``could not acquire lock'' & 6--6 & 6--6\\
Succeeded during the outage & 0--0 & 0--0\\
Failed but committed & 1--1 & 1--1\\
Runs with health 200 throughout & 5 of 5 & 5 of 5\\
Runs with all tasks alive; postings back after & 5; 5 of 5 & 5; 5 of 5\\}
\newcommand{\KillPostOld}{57.9\,s [55.6, 66.4]}
\newcommand{\KillPostNew}{17.2\,s [16.2, 22.1]}
\newcommand{\KillOtherNew}{17.2\,s [16.3, 22.0]}

\newcommand{\FailSection}{%
Both failure scenarios were repeated on each edition (\FailReps{} repetitions), each on a fresh deployment from the seed
snapshot with the workload's accounts, editions alternating as in the performance rounds. Times are measured in the
emulation; Section~\ref{sec:threats} discusses how they would differ on AWS.

\textbf{Task crash under load (E3).} Eight clients posted continuously for 30\,s. At 8\,s the task running the
posting engine was killed with SIGKILL (the transactions service in the nine-service edition, the API host in the
edition) and a replacement was started 3\,s later, as ECS would do. A prober read one endpoint of each module every
100\,ms. Table~\ref{tab:kill} gives the results.

\begin{table}[t]
\centering\footnotesize
\caption{Killing the posting task under load, emulated: median [min, max] over \FailReps{} repetitions, or totals.}
\label{tab:kill}
\setlength{\tabcolsep}{3pt}
\begin{tabularx}{\columnwidth}{@{}Lrr@{}}
\toprule
& Nine-service & Edition\\
\midrule
Task killed & transactions & API host\\
\FailRows
\bottomrule
\end{tabularx}
\\[2pt]{\scriptsize $^c$ nine-service edition: the customers service's first request after an idle period took
about 13\,s at 0.25 vCPU while it built its data model, so the prober's 5\,s requests timed out; the transactions
crash did not cause it.}
\end{table}

Neither edition lost an acknowledged posting or wrote a partial transaction in any repetition, and every ledger
passed every check afterwards. In some runs a posting was committed whose client saw an error: the window of an
in-flight request, after which the transaction shows as posted. Composition changes two things, in opposite directions. The first is the failure
domain: in the nine-service edition accounts and users kept answering while the transactions service was down,
whereas in the edition every module was unavailable (worst other module \KillOtherNew). Customers reads in the
nine-service edition also failed, but the service logs show the cause was a slow first request after idle, not the
crash (Table~\ref{tab:kill}). The second is recovery:
the replacement transactions service, with a quarter of a vCPU, starts and warms up slowly under load, so postings
were unavailable for \KillPostOld, against \KillPostNew{} for the edition's API host with a full vCPU.

\textbf{Projection to AWS (E5).} Unavailability exceeds restart time in both editions because the emulated ALB, like
a real one, keeps a failed target out for a while. On AWS the edition's target group needs two passing health
checks 30\,s apart before a replaced task receives traffic, so absolute outages would be at least a minute longer in
both editions. We expect the difference in failure domain to carry over, since it follows from the topology; we did
not measure it on AWS.

\textbf{AWS endpoint outage (E3).} The emulated AWS endpoint was frozen for 45\,s, so calls hang as in a network
partition, while eight postings were attempted (Table~\ref{tab:outage}). The posting path publishes to SNS while it
holds the account lock, so one hung publish blocks every posting to the same accounts until the client gives up,
and the rest are rejected after about 3\,s. Some failed postings were nevertheless committed. Health endpoints
answered 200 throughout, all tasks survived, and postings succeeded again once the endpoint returned. The pattern is
the same in both editions: composition neither causes nor cures it.

\begin{table}[t]
\centering\footnotesize
\caption{Eight postings during a 45\,s freeze of the emulated AWS endpoint: range over \FailReps{} repetitions.}
\label{tab:outage}
\begin{tabularx}{\columnwidth}{@{}Lrr@{}}
\toprule
& Nine-service & Edition\\
\midrule
\OutageRows
\bottomrule
\end{tabularx}
\end{table}

\textbf{Answer to RQ3 (failure).} In the emulation both editions keep the ledger consistent through repeated hard
crashes and an AWS stall. Composition enlarges the failure domain of a crash to all five modules and, at these task
sizes, shortens recovery of the posting path.%
}

\begin{document}

\title{A Consolidated Deployment Edition of a Microservice Core-Banking Platform:\\
Composing Services by Reference into Three Hosts}

\author{\IEEEauthorblockN{Tomisin Fagbola}
\IEEEauthorblockA{SuperSolutions Technology Limited}}

\maketitle

\begin{abstract}
Microservices give teams independent deployment and scaling, but each deployed service also reserves compute and
adds operational work. This paper examines a second deployment edition of a multi-tenant .NET core-banking platform.
The existing edition runs as nine AWS ECS Fargate tasks; the alternative composes the five API services by project
reference into one API host, alongside a consumer host and a background host. The service code remains in its
original repositories, and both editions continue to be built from the same commits. We describe the composition
mechanism, module manifest, start-up checks, request handling and configuration precedence, then compare the editions
in an emulation of the AWS deployment. The composed edition exposes all 844 declared operations. Across 47 financial
workload cases, both editions produced the same observed outcomes and matching ledgers. In read probes across five
principal types, 230 of 231 endpoint results matched; the remaining difference was associated with the emulated load
balancer. Eleven cross-tenant and write-authorisation cases also produced the expected outcomes in both editions. In
the emulation, the composed edition delivered 2.3 times the read throughput and used less than half the memory.
Estimated service-task compute at list prices was 11--24\% lower, excluding the shared gateway, load balancer, data
stores and other operating costs. The consolidation also enlarges the failure domain: one API-host crash takes all
five modules offline. Of 16 injected faults, the edition's checks detected 10, behavioural tests detected five
additional faults, and one fault escaped both. These findings concern one platform and an emulated deployment that
mirrors its topology, task sizes and configuration; they are not production-AWS measurements.
\end{abstract}

\begin{IEEEkeywords}
modular monolith, service composition, core banking, differential testing, fault injection, AWS Fargate,
experience report
\end{IEEEkeywords}

% =====================================================================================================
\section{Introduction}

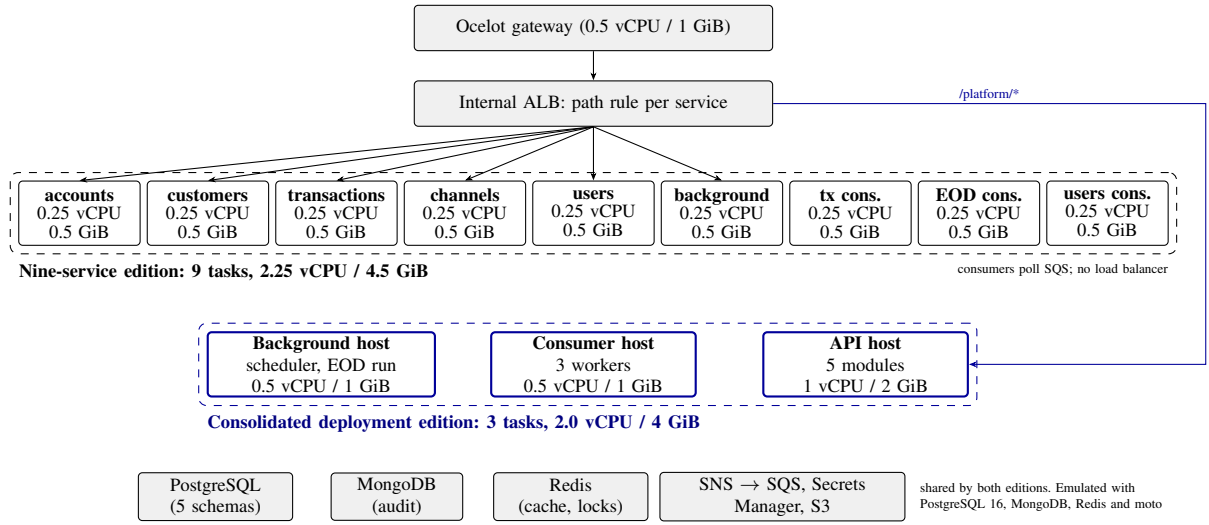
\begin{figure*}[t]
\centering
\begin{tikzpicture}[
  font=\scriptsize,
  task/.style={draw, rounded corners=1.5pt, align=center, inner sep=1.5pt, minimum height=0.85cm, text width=#1},
  task/.default=1.5cm,
  infra/.style={draw, fill=gray!12, rounded corners=1.5pt, align=center, inner sep=2pt, minimum height=0.6cm},
  arr/.style={-{Stealth[length=3pt]}, thin},
  grp/.style={draw, dashed, rounded corners=3pt, inner sep=3pt},
  lbl/.style={font=\tiny, fill=white, inner sep=1pt}]
  \node[infra, text width=4.6cm] (gw) at (0,0.1) {Ocelot gateway (0.5 vCPU / 1 GiB)};
  \node[infra, text width=4.6cm] (alb) at (0,-0.9) {Internal ALB: path rule per service};
  \draw[arr] (gw) -- (alb);
  % nine-service edition
  \foreach \n/\t/\i in {acc/accounts/0, cus/customers/1, trx/transactions/2, chn/channels/3, usr/users/4, bg/background/5,
                        txc/tx cons./6, eod/EOD cons./7, uc/users cons./8} {
     \node[task] (\n) at ({-6.8+\i*1.7},-2.35) {\mbox{\textbf{\t}}\\[-1pt]0.25 vCPU\\[-1pt]0.5 GiB};
  }
  \foreach \n in {acc,cus,trx,chn,usr,bg} \draw[arr] (alb.south) -- (\n.north);
  \node[grp, fit=(acc)(uc)] (oldg) {};
  \node[font=\scriptsize\bfseries, anchor=north west] at (oldg.south west) {Nine-service edition: 9 tasks, 2.25 vCPU / 4.5 GiB};
  \node[font=\tiny, anchor=north east, align=right] at (oldg.south east) {consumers poll SQS; no load balancer};
  % consolidated deployment edition
  \node[task=2.9cm, draw=blue!60!black, thick] (bgs) at (-3.6,-4.35) {\textbf{Background host}\\scheduler, EOD run\\0.5 vCPU / 1 GiB};
  \node[task=2.6cm, draw=blue!60!black, thick] (con) at (0.0,-4.35) {\textbf{Consumer host}\\3 workers\\0.5 vCPU / 1 GiB};
  \node[task=2.6cm, draw=blue!60!black, thick] (api) at (3.6,-4.35) {\textbf{API host}\\5 modules\\1 vCPU / 2 GiB};
  \node[grp, draw=blue!60!black, fit=(bgs)(api)] (newg) {};
  \node[font=\scriptsize\bfseries, anchor=north west, text=blue!50!black] at (newg.south west) {Consolidated deployment edition: 3 tasks, 2.0 vCPU / 4 GiB};
  \draw[arr, blue!60!black] (alb.east) -- node[lbl, above]{/platform/*} (8.1,-0.9) |- (api.east);
  % stores
  \node[infra, text width=1.9cm] (pg) at (-5.0,-6.1) {PostgreSQL\\(5 schemas)};
  \node[infra, text width=1.6cm] (mg) at (-2.6,-6.1) {MongoDB\\(audit)};
  \node[infra, text width=1.9cm] (rd) at (-0.3,-6.1) {Redis\\(cache, locks)};
  \node[infra, text width=3.1cm] (aws) at (2.5,-6.1) {SNS $\rightarrow$ SQS, Secrets\\Manager, S3};
  \node[font=\tiny, align=left, anchor=west] at (4.2,-6.1) {shared by both editions. Emulated with\\PostgreSQL 16, MongoDB, Redis and moto};
\end{tikzpicture}
\caption{The two editions as deployed on ECS Fargate in eu-west-1, with task sizes from the staging estate and from the
edition's task definitions, and as emulated in this study. Both sit behind the same gateway and internal ALB and
share one set of stores. Allocated compute, excluding the gateway, is 2.25 vCPU / 4.5 GiB for the nine tasks against
2.0 vCPU / 4 GiB for the three.}
\label{fig:topology}
\end{figure*}

The platform studied here is a multi-tenant core-banking system written in .NET. Its capabilities (customers,
channels, transactions, accounts and users, with consumers and a scheduler) were built as separate services in
seven repositories, and it ships as nine ECS Fargate tasks behind an Ocelot gateway and an internal Application
Load Balancer (ALB). Some customers asked for fewer moving parts. The answer is a second edition of the same
product: one repository whose API host references the five API services' projects and serves their controllers as
MVC application parts, beside a consumer host for the three queue workers and a background host for the scheduler
(Fig.~\ref{fig:topology}). The nine-service edition is not retired; both editions ship from the same service
repositories, so every fix made in a service reaches both.

An earlier version of this report described the composition and its checks but could not say whether the edition
behaves like the services it composes, what it costs, or how it fails. This version keeps that architectural
description, updated where the code has moved (Sections~\ref{sec:drivers}--\ref{sec:database}), and answers those
questions by experiment. Both editions were built from the same commits and run side by side in an emulation of their AWS
deployment, and compared differentially \cite{mckeeman1998differential}: the same writes, the same reads, the same
failures and the same load.

\textbf{Terms.} We keep three claims apart throughout. \term{Route parity} means that every operation a standalone
service declares is served by the composed host at the same method and path. \term{Tested behavioural equivalence}
means that the requests we sent produced matching outcomes, after a stated normalisation, in both editions.
\term{System-wide equivalence}, that every possible request behaves alike, is not established by this study.
We also keep four questions apart: whether the two editions behaved alike (\term{equivalence}), whether the behaviour
followed the bank's intended rules (\term{compliance}), whether each ledger stayed balanced (\term{consistency}), and
whether each response matched what the test script expected. Passing one does not imply passing the others;
Section~\ref{sec:rq1} reports all four.

\textbf{Contributions.} The paper makes four contributions.
\begin{enumerate}
\item \emph{A method} for composing independently developed services into three hosts (one API host for the five API
services, plus a consumer host and a background host) without copying their code, so both editions keep shipping
from the same repositories (Sections~\ref{sec:drivers}--\ref{sec:database}).
\item \emph{A side-by-side comparison} of the two editions on emulated AWS, covering behaviour, failures, performance
and cost (Sections~\ref{sec:rq1}, \ref{sec:rq3fail} and~\ref{sec:rq3perf}).
\item \emph{A test of the safety checks}: which of 16 injected faults the edition's own checks catch, and which only
behavioural tests catch (Section~\ref{sec:rq2}).
\item \emph{The failure modes met along the way}, and when they matter (Section~\ref{sec:findings}).
\end{enumerate}
Service composition, modular monoliths and deployment consolidation are not new. What this study adds is a measured,
differential evaluation in which both editions are the same production platform, built from the same commits:
the services are composed by reference and keep shipping on their own, so the composed edition can be compared with
the deployment it would replace, at its real task sizes, and judged against intended behaviour as well as against
the other edition.

% =====================================================================================================
\section{Background and Related Work}
\label{sec:related}

\textbf{Core banking.} A core-banking system is a bank's system of record. It holds the institution's customers,
their accounts and balances, and the transactions that change those balances, and it applies the rules by which money
moves: a debit has a matching credit, an account may not be overdrawn beyond its limit, and a posting belongs to a
business day. Mobile and web applications, agent terminals and cards are channels that call the core; payment rails,
fraud monitoring and reporting are neighbouring systems. Four groups of capability recur: customer and account
management; transaction processing, which validates transactions and posts them to accounts and a general ledger;
end-of-day processing, which closes a business day, computes balances and charges and opens the next, usually as
batch work; and identity and access. The platform here has the same shape: modules for customers, channels,
transactions, accounts and users, consumers for transactions, end-of-day work and user events, and a scheduler. Core
banking also has requirements most web applications do not: money must be neither created nor lost, so postings need
transactional guarantees; changes should be auditable and are often made by one user and approved by another
(maker-checker); and a platform that serves several institutions must keep each one's data apart, which this one does
by tenant, with a global query filter in a shared library.

\begin{figure}[t]
\centering\begin{tikzpicture}[node distance=2mm]

% A. Monolith
\node[ttl, anchor=west] at (-3.75,0.55) {A. Monolith};
\node[fb=1.6cm, cB] (ao) at (-2.3,-0.25) {Customers};
\node[fb=1.6cm, cB] (aa) at (0,-0.25) {Accounts};
\node[fb=1.6cm, cB] (ap) at (2.3,-0.25) {Payments};
\draw[ar, <->] (ao) -- (aa); \draw[ar, <->] (aa) -- (ap);
\node[grpbox, fit=(ao)(ap), inner ysep=7pt, label={[lab]above:one deployable unit}] (ag) {};
\node[fb=2.2cm, cN, below=4mm of ag.south west, anchor=north west, xshift=0.6cm] (adb) {one database};
\draw[ar] (ag.south -| adb.north) -- (adb.north);
\node[lab, anchor=east] at (3.75,-1.25) {any module may call any other};
% B. Microservices
\node[ttl, anchor=west] at (-3.75,-1.85) {B. Microservices};
\node[lab] at (0,-1.85) {calls over the network};
\foreach \n/\x in {Customers/-2.3, Accounts/0, Payments/2.3} {
  \node[fb=1.6cm, cB] (b\n) at (\x,-2.45) {\n};
  \node[grpbox, fit=(b\n), inner ysep=5pt, label={[lab]below:own unit}] (bg\n) {};
  \node[fb=1.6cm, cN, font=\tiny] (bd\n) at (\x,-3.65) {own database};
  \draw[ar] (bg\n.south) ++(0,-0.27) -- (bd\n.north);
}
\draw[ar] (bgCustomers) -- (bgAccounts); \draw[ar] (bgAccounts) -- (bgPayments);
\node[lab, anchor=east] at (3.75,-4.15) {each service is deployed and scaled alone};
% C. Modular monolith
\node[ttl, anchor=west] at (-3.75,-4.6) {C. Modular monolith};
\foreach \n/\x in {Customers/-2.3, Accounts/0, Payments/2.3} {
  \node[fb=1.6cm, cB, minimum height=0.65cm] (c\n) at (\x,-5.45) {\n};
  \node[draw=black!70, fill=green!13, minimum width=0.6cm, minimum height=0.13cm, inner sep=0pt] at (c\n.north) {};
}
\node[lab] at (-2.3,-4.95) {interface};
\draw[ar] (cCustomers) -- (cAccounts); \draw[ar] (cAccounts) -- (cPayments);
\node[grpbox, fit=(cCustomers)(cPayments), inner ysep=9pt, label={[lab]above right:one deployable unit}] (cg) {};
\node[fb=5.8cm, cN, below=4mm of cg] (cdb) {one database (strict form: a schema per module)};
\draw[ar] (cg) -- (cdb);
\node[lab, below=1.5mm of cdb.south east, anchor=north east] {modules call each other only through their interfaces};

\end{tikzpicture}
\caption{Three styles: a monolith (one unit, any module may call any other), microservices (one unit and one store per
service, calls over the network) and a modular monolith (one unit, modules that meet only through interfaces).}
\label{fig:styles}
\end{figure}
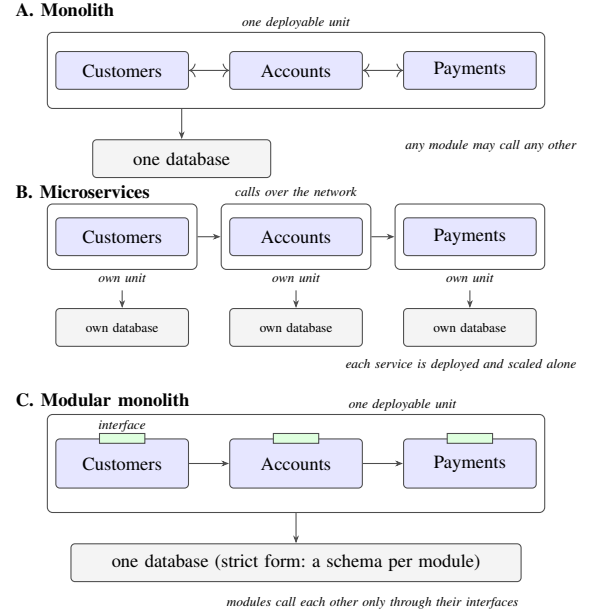

\textbf{Styles.} A monolith is one deployable unit; microservices are independently deployable services that each
own a capability \cite{lewis2014microservices,newman2015building,dragoni2017microservices}; a modular monolith keeps
one unit but divides it into modules with explicit boundaries
\cite{fowler2015monolithfirst,westeinde2019deconstructing,newman2019monolith}, an idea that goes back to information
hiding \cite{parnas1972criteria} and bounded contexts \cite{evans2003ddd}. A grey-literature review by Su and Li
\cite{su2024modular} finds the modular monolith presented in industry both as an alternative to microservices and as
a step before them. Fig.~\ref{fig:styles} and Table~\ref{tab:styles} compare the styles. The edition is a \term{deployment-level modular
monolith} for the API tier: one process serves five modules behind a common contract, registered explicitly and
checked at start-up, but it lacks the strict form's isolation. The modules use the shared library's data contexts,
which the host owns, against one database; configuration is layered into one flat namespace, so one module's value for
a key replaces another's (Section~\ref{sec:config}); a filter registered globally by one module would apply to all; and
one process can load only one version of each dependency (Section~\ref{sec:deps}). The edition is also three hosts,
not one process. We use \term{modular monolith} for the general style, \term{deployment-level modular monolith} for the
API tier of this edition, and \term{consolidated deployment edition}, or \term{the edition}, for the product variant.

\begin{table}[t]
\centering\footnotesize
\caption{The styles compared, and the edition described in this paper.}
\label{tab:styles}
\begin{tabularx}{\columnwidth}{@{}lLLL@{}}
\toprule
Style & Deployable units & Module boundaries & Data\\
\midrule
Monolith & one & not enforced by structure & one shared store\\
Microservices & many & network boundaries & typically one store per service\\
Modular monolith (strict) & one & explicit module interfaces & typically one schema per module\\
This edition & three hosts; five API modules in one & convention and start-up checks & shared contexts, one database\\
\bottomrule
\end{tabularx}
\end{table}

\textbf{Vocabulary.} We use the vocabulary of software architecture practice \cite{bass2021software}. A
quality-attribute scenario states a stimulus, the environment it arrives in and the wanted response; quality attributes
are named as in ISO/IEC 25010 \cite{iso25010}; the architecture is described through views \cite{kruchten1995views,c4model};
choices are recorded as decisions with their alternatives and consequences \cite{nygard2011adr}; and, following the
architecture trade-off analysis method \cite{kazman1998atam}, we separate sensitivity points, trade-off points and
risks.

\begin{table*}[t]
\centering\footnotesize
\caption{Closest related work, compared on what each reports. (a) real production system; (b) both styles built
and run; (c) behavioural equivalence tested; (d) failure injected; (e) cost measured or modelled; (f) checks that
keep two deployments in step evaluated. \yes: reported; \no: not reported.}
\label{tab:related}
\begin{tabularx}{\textwidth}{@{}p{4.3cm}p{2.9cm}Lcccccc@{}}
\toprule
Work & Setting & Direction & (a) & (b) & (c) & (d) & (e) & (f)\\
\midrule
Villamizar et al.\ \cite{villamizar2015evaluating,villamizar2016cost} & web app on AWS & monolith vs.\ microservices & \no & \yes & \no & \no & \yes & \no\\
Blinowski et al.\ \cite{blinowski2022monolithic} & benchmark app & monolith vs.\ microservices & \no & \yes & \no & \no & \no & \no\\
Salunkhe \cite{salunkhe2022load} & five-service app & monolith vs.\ microservices & \no & \yes & \no & \no & qual. & \no\\
Aydemir and Ba\c{s}\c{c}ift\c{c}i \cite{aydemir2022corebanking} & commercial bank & monolith $\rightarrow$ microservices & \yes & \yes & \no & \no & \no & \no\\
Fritzsch et al.\ \cite{fritzsch2019monolith}; Jamshidi et al.\ \cite{jamshidi2018microservices} & surveys & monolith $\rightarrow$ microservices & \no & \no & \no & \no & \no & \no\\
Su, Li and Taibi \cite{su2023back} & multivocal review, 4 cases & microservices $\rightarrow$ monolith & \yes & \no & \no & \no & reported & \no\\
Service Weaver \cite{ghemawat2023modern,johnson2024weaver} & framework & one codebase, deployed either way & \no & \yes & \no & \no & \yes & \no\\
\textbf{This paper} & multi-tenant core banking & existing services composed by reference; both editions ship & \yes & \yes & \yes & \yes & \yes & \yes\\
\bottomrule
\end{tabularx}
\end{table*}

\textbf{Related work.} Table~\ref{tab:related} places this report. Comparisons of monolithic and microservice
deployments \cite{villamizar2015evaluating,villamizar2016cost,blinowski2022monolithic,salunkhe2022load} build an
application twice in order to measure it, and migration studies \cite{jamshidi2018microservices,fritzsch2019monolith}
describe moving from one style to the other; the closest setting is the core-banking system of Aydemir and
Ba\c{s}\c{c}ift\c{c}i \cite{aydemir2022corebanking}, which moved to microservices. Two lines of work come closer to
our direction. Su, Li and Taibi \cite{su2023back} review cases of organisations that moved from microservices back
to a monolith, including Prime Video and Segment, chiefly for cost; those moves replace the microservices, whereas
here both editions keep shipping. Service Weaver \cite{ghemawat2023modern} lets an application be written as a
modular monolith and deployed either as one process or as microservices, and an independent evaluation finds
separate code bases among the features it lacks \cite{johnson2024weaver}. Our case is the inverse starting point:
code bases that were already separate, owned and released independently, composed after the fact without moving
their code. A targeted search of arXiv, DBLP and the web in October 2026 (terms: \emph{modular monolith},
\emph{microservice to monolith}, \emph{microservices consolidation}, \emph{Service Weaver}, \emph{composing
microservices into one host}) found no study that composes existing services into a second edition shipped beside
them and tests the two for equivalence. The search was not systematic, so we claim an under-reported combination,
not a first. Our evaluation borrows two established techniques: differential testing, which judges a system by
comparing it with another implementation of the same behaviour \cite{mckeeman1998differential}, and fault injection
in the spirit of mutation analysis \cite{jia2011mutation}, which measures a test suite by the faults it detects.

% =====================================================================================================
\section{The Platform and the Two Editions}
\label{sec:platform}

Both editions use PostgreSQL (five schemas, one per shared data context), MongoDB for audit records, Redis for
cache and posting locks, SNS topics fanned out to SQS queues for asynchronous work, Secrets Manager for
configuration, and S3. All services depend on one shared library, distributed as packages from a private feed,
which carries the data contexts, the tenant filter, caching, messaging and the posting engine.

\textbf{The cost problem.} Nine services meant nine deployments (Fig.~\ref{fig:reference}), each with its own compute
allocation, load-balancer routing, queue consumers, database connections and telemetry. For a small deployment almost all of that capacity was idle, so the operating cost reflected the number of services rather
than the amount of work. A container service bills for what a task reserves, not for what it uses, so the allocation,
not the load, sets a floor on the cost. The earlier report argued this qualitatively; Section~\ref{sec:rq3perf}
estimates it. The constraint the team set was that the fix must not change what a client sees: a client moves to the
new deployment by changing only its base address, and every module keeps its code and repository.

\begin{figure}[t]
\centering\begin{tikzpicture}[node distance=2mm]

\node[fb=1.5cm, cW] (ch) at (-3,0) {Channel apps};
\node[fb=1.6cm, cB] (gw) at (-0.6,0) {Gateway / firewall};
\node[fb=1.5cm, cB] (lb) at (1.8,0) {Load balancer};
\draw[ar] (ch) -- (gw); \draw[ar] (gw) -- (lb);
\node[font=\tiny\bfseries] at (-1.35,-0.95) {API services (container tasks)};
\foreach \t/\x/\y/\c in {Customer/-2.6/-1.35/cY, Channel/-1.35/-1.35/cY, Transaction/-0.1/-1.35/cY,
                         Account/-2.6/-1.85/cY, User/-1.35/-1.85/cY, {Fixed assets}/-0.1/-1.85/cN} {
  \node[fb=1.12cm, \c, minimum height=0.38cm, font=\tiny, inner sep=1pt] at (\x,\y) {\t};
}
\node[draw=black!70, inner sep=3pt, fit={(-3.15,-0.75) (0.45,-2.1)}] (api) {};
\node[font=\tiny\bfseries] at (2.25,-0.95) {Workers};
\foreach \t/\y in {Transaction consumer/-1.3, EOD consumer/-1.7, Background services/-2.1} {
  \node[fb=2.1cm, cG, minimum height=0.32cm, font=\tiny] at (2.25,\y) {\t};
}
\node[draw=black!70, inner sep=3pt, fit={(1.1,-0.75) (3.4,-2.3)}] (wk) {};
\node[fb=2.0cm, cR] (mq) at (2.45,-2.85) {Message queue};
\draw[ar] (wk) -- (mq); \draw[ar, dashed] (api.east) ++(0,-0.45) -- (mq.west);
\draw[ar] (lb) -- (1.8,-0.62);
\node[db, fill=violet!13] (rdb) at (-2.4,-3.95) {Relational database};
\node[db, fill=green!13] (cache) at (-0.6,-3.95) {Cache};
\node[db, fill=yellow!25] (obj) at (1.2,-3.95) {Object store};
\draw[ar] (rdb.north |- api.south) -- (rdb.north); \draw[ar] (cache.north |- api.south) -- (cache.north);
\node[draw=black!60, dashed, inner sep=4pt, fit=(api)(wk)(mq)(rdb)(obj), label={[lab, anchor=south west]north west:Private network}] {};
\node[fb=7.7cm, cW, anchor=north, text width=7.7cm, align=left] at (0.1,-4.75)
  {\textbf{Cost behaviour.} \emph{Repeated for every deployed service:} compute allocation, load-balancer routing rule,
   queue consumer, database connections, log and metric streams. \emph{Paid once whatever the number of services:}
   load-balancer base charge, database instance, cache, queue.};

\end{tikzpicture}
\caption{Generic reference deployment before composition, and which costs repeat per service and which are paid once.}
\label{fig:reference}
\end{figure}

Table~\ref{tab:tasks} lists the tasks as deployed in eu-west-1. The nine-service edition runs every service at the
smallest Fargate size, 0.25 vCPU and 0.5 GiB; the edition's task definitions give the API host 1 vCPU and 2 GiB and
the other two hosts 0.5 vCPU and 1 GiB. The gateway (0.5 vCPU, 1 GiB) and the ALB are shared. The edition's
Terraform registers the API host as one target group under \code{/platform/*} with a readiness check every 30\,s and
two passes to become healthy.

\begin{table}[t]
\centering\footnotesize
\caption{Fargate tasks of each edition (eu-west-1, Linux/x86, one task each).}
\label{tab:tasks}
\begin{tabularx}{\columnwidth}{@{}Lccc@{}}
\toprule
Task & vCPU & Memory (GiB) & Behind ALB\\
\midrule
\multicolumn{4}{@{}l}{\emph{Nine-service edition}}\\
accounts, customers, transactions, channels, users, background (6) & 0.25 each & 0.5 each & \yes\\
transaction, end-of-day and users consumers (3) & 0.25 each & 0.5 each & \no\\
\emph{Total} & \emph{2.25} & \emph{4.5} & 6 targets\\
\midrule
\multicolumn{4}{@{}l}{\emph{Consolidated deployment edition}}\\
API host (5 modules) & 1.0 & 2 & \yes\\
Consumer host (3 workers) & 0.5 & 1 & \no\\
Background host & 0.5 & 1 & \no\\
\emph{Total} & \emph{2.0} & \emph{4} & 1 target\\
\midrule
Gateway (shared) & 0.5 & 1 & --\\
\bottomrule
\end{tabularx}
\end{table}

% =====================================================================================================
\section{Architectural Drivers and Constraints}
\label{sec:drivers}

The edition exists because of a few quality attributes \cite{bass2021software,iso25010}. Table~\ref{tab:qas} states
each as a quality-attribute scenario, with the response recorded in the repository and, where this study measured
it, the measured response. The scenarios are our wording of goals the repository states or implies, not
requirements that were recorded in this form.

\begin{table*}[t]
\centering\footnotesize
\caption{Quality-attribute scenarios that drive the edition, with the response as designed and as measured here.}
\label{tab:qas}
\begin{tabularx}{\textwidth}{@{}lLLL@{}}
\toprule
Attribute & Scenario & Response as designed & Response measured in this study\\
\midrule
Operability, cost & A customer wants fewer moving parts than nine services & Three hosts & 3 tasks instead of 9; 11--24\% estimated saving in service-task compute, shared infrastructure excluded (Section~\ref{sec:rq3perf})\\
Compatibility & A client changes only its base URL & Every operation preserved & 844 of 844 operations; tested behavioural equivalence (Section~\ref{sec:rq1})\\
Modifiability & A fix is made in one service & It appears in both editions, no second edit & 18 guarded corrections; the accounts service's current 288 operations (282 in the earlier report) reached both editions unedited\\
Data compatibility & The edition runs on an existing database & No schema change, by construction; migrations behind a flag & Both editions ran on one schema and produced identical ledgers\\
Availability & A background worker throws & The host stays up & Confirmed; health stays green, so the failure is quiet (P3)\\
Security & A module adds a global filter or middleware & Module-scoped filters stay within their modules & 87 and 369 actions; a leak stops start-up (F05), a dropped filter does not (F06)\\
Verifiability & Composition breaks something no build catches & A check per recorded failure & 10 of 16 injected faults caught by the checks (Section~\ref{sec:rq2})\\
\bottomrule
\end{tabularx}
\end{table*}

\textbf{Constraints.} The edition does not own the service code: the seven repositories keep their own pipelines and
deployments. Modules are pre-compiled against the dependency versions their own repository chose, and a process
loads one version of each dependency. The existing gateway and shared stores are kept, the schema may not change,
and no credential may sit in a container layer or a configuration file; the private feed is reached through an
environment variable. Most of the design follows from these constraints, and the compatibility and modifiability
scenarios together explain why composition is by reference.

% =====================================================================================================
\section{Composition}
\label{sec:composition}

\subsection{Architectural views}
\begin{figure}[t]
\centering\begin{tikzpicture}[node distance=2mm]

\node[fb=3.2cm, cN] (cl) at (-2.1,0) {Client applications};
\node[fb=3.2cm, cN] (gw) at (2.1,0) {Existing gateway};
\draw[ar] (cl) -- (gw);
\node[fb=3.9cm, cB] (api) at (-1.6,-1.45) {API host: five modules};
\node[fb=3.9cm, cB, below=1.5mm of api] (bg) {Background host};
\node[fb=3.9cm, cG, below=1.5mm of bg] (con) {Consumer host (polls queues)};
\node[grpbox, fit=(api)(con), inner sep=4pt, label={[ttl]above:Consolidated deployment edition}] (ed) {};
\draw[ar] (gw.south) -- (ed.north east);
\node[fb=2.6cm, cY] (id) at (2.75,-1.45) {Identity: the users service issues tokens; the provider's realm for channels};
\node[fb=2.6cm, cN] (ext) at (2.75,-2.6) {External payment and notification services};
\draw[ar] (ed.east |- id) -- (id); \draw[ar] (ed.east |- ext) -- (ext);
\foreach \n/\t/\x in {pg/PostgreSQL/-3.0, mg/{MongoDB (audit)}/-1.0, rd/{Redis (cache)}/1.0, sq/{SNS, SQS (queues)}/3.0} {
  \node[fb=1.75cm, cN] (\n) at (\x,-3.85) {\t};
  \draw[ar] (ed.south) -- (\n.north);
}

\end{tikzpicture}
\caption{Context view: clients, the existing gateway, the three hosts, identity, and the stores they use.}
\label{fig:context}
\end{figure}
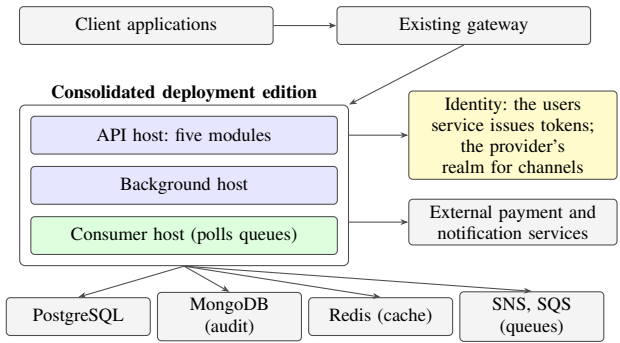

We describe the edition through the views of Kruchten's 4+1 model \cite{kruchten1995views} and a context view in the
style of the C4 model \cite{c4model} (Table~\ref{tab:views}). The context view (Fig.~\ref{fig:context}) shows the
edition as one unit between the existing gateway and the stores it uses. Identity is split: the users module issues
tokens, while the channels module validates tokens from an external identity provider's realm. This is a description
of the architecture as built, not a formal ATAM evaluation \cite{kazman1998atam}, which would involve stakeholders.

\begin{table}[t]
\centering\footnotesize
\caption{The views used to describe the edition.}
\label{tab:views}
\begin{tabularx}{\columnwidth}{@{}lLl@{}}
\toprule
View & Concern & Shown in\\
\midrule
Context & the edition, its users, the systems around it & Fig.~\ref{fig:context}\\
Logical & modules, their contract, host-owned concerns & Figs.~\ref{fig:apihost}--\ref{fig:scoped}\\
Process & hosts, start-up and failure behaviour & Figs.~\ref{fig:startup}, \ref{fig:consumer}, \ref{fig:background}; Table~\ref{tab:startdb}\\
Development & sibling repositories, pins, configuration layers & Figs.~\ref{fig:ceiling}--\ref{fig:flatkey}; Listing~\ref{lst:layout}\\
Physical & tasks, load balancers, gateway & Fig.~\ref{fig:topology}; Table~\ref{tab:tasks}\\
Scenarios & the drivers that tie the views together & Table~\ref{tab:qas}\\
\bottomrule
\end{tabularx}
\end{table}

\subsection{Composition by reference}
The edition's repository contains no service code. Its API host references each service's API project by relative
path, so the eight checkouts must sit side by side (Listing~\ref{lst:layout}). A layout script checks this and says
what is missing, and in continuous integration the same layout is assembled by a multi-repository checkout, so the
build does not depend on any one machine. Each service is built from its own mainline: five repositories use a branch named
\code{development} and two use \code{dev}, which is why the pipeline takes two branch inputs and the
environment-check script expects different branches for different repositories. Each service exposes a
registration extension, added in its own repository as a pure refactor where its wiring was inline and verified by
its own tests, and the host calls it. This differs from merging the services into one application: nothing moves,
each service keeps its repository, pipeline, tests and standalone deployment, and a fix made in a service reaches
both editions with no second edit.

\begin{figure}[t]
\begin{lstlisting}[float=false,caption={Required layout of the sibling checkouts.},label={lst:layout}]
<some parent>/
  CoreBanking.Platform/              <- the edition's repository
  CoreBanking.AccountManagement/
  CoreBanking.Channels.Middleware/
  CoreBanking.CustomerManagement/
  CoreBanking.TransactionApi/
  CoreBanking.UserManagement/
  CoreBanking.TransactionConsumer/
  CoreBanking.BackgroundServices/
\end{lstlisting}
\end{figure}

\subsection{Modules and the API host}
\begin{figure}[t]
\centering\begin{tikzpicture}[node distance=2mm]

\foreach \m/\x in {Customers/-3.2, Channels/-1.6, Transactions/0, Accounts/1.6, Users/3.2} {
  \node[fb=1.25cm, cB, font=\tiny] (m\m) at (\x,-0.9) {\m\\ module\\ (IModule)};
  \node[fb=1.25cm, cN, font=\tiny] (s\m) at (\x,-2.55) {service\\ project\\ (own repo)};
  \draw[ar] (m\m) -- (s\m);
}
\node[font=\scriptsize, align=center] (ht) at (0,0.15) {API host: logging, telemetry, health, authentication,\\ shared data contexts, audit sink, cache, module registry};
\node[grpbox, fit=(ht)(mCustomers)(mUsers)] {};
\node[lab, fill=white, align=center] at (0,-1.75) {controllers served as MVC application parts;\\ each service's own registration is called};

\end{tikzpicture}
\caption{The API host and its five modules. Each module references a service project in its own repository.}
\label{fig:apihost}
\end{figure}
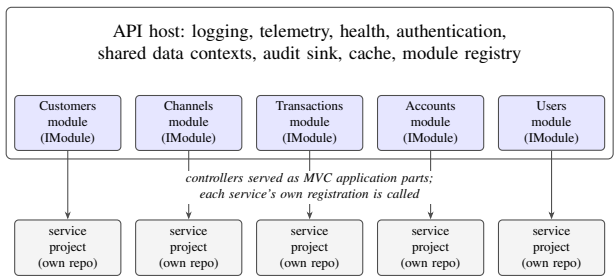

The API host composes five modules: customers, channels, transactions, accounts and users (Fig.~\ref{fig:apihost}).
A module is registered explicitly in the host's start-up file, not discovered by scanning, so the file is the
manifest of what the edition contains and a missing module is visible instead of silent. To add a module, one adds a
project reference to the service's API project, an implementation of the module contract that names the module's
controller assembly, route prefix, middleware, scoped filters and registration extension, and an entry in the modules
array. Registration is one call (Listing~\ref{lst:register}).

\begin{figure}[t]
\begin{lstlisting}[float=false,language={[Sharp]C},caption={Registering modules; validation and the scoped-filter convention are applied here (condensed).},label={lst:register}]
public static IServiceCollection AddModules(
    this IServiceCollection services,
    IConfiguration config, params IModule[] modules)
{
    Validate(modules);              // throws on a bad composition
    var mvc = services.AddControllers();
    foreach (var m in modules)
    {
        foreach (var asm in m.ControllerAssemblies)
            mvc.AddApplicationPart(asm);
        m.RegisterServices(services, config);
    }
    if (modules.Any(m => m.ScopedFilters.Count > 0))
    {
        var scope =
            new ModuleScopedFilterConvention(modules);
        services.Configure<MvcOptions>(
            o => o.Conventions.Add(scope));
    }
    services.AddSingleton<IReadOnlyCollection<IModule>>(
        modules);
    return services;
}
\end{lstlisting}
\end{figure}

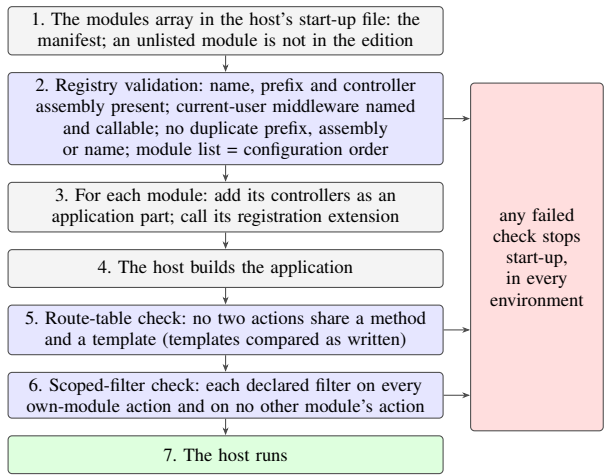
\begin{figure}[t]
\centering\begin{tikzpicture}[node distance=2mm]
\node[fb=5.6cm, cN] (s1) {1. The modules array in the host's start-up file: the manifest; an unlisted module is not in the edition};
\node[fb=5.6cm, cB, below=2.2mm of s1] (s2) {2. Registry validation: name, prefix and controller assembly present; current-user middleware named and callable; no duplicate prefix, assembly or name; module list = configuration order};
\draw[ar] (s1) -- (s2);
\node[fb=5.6cm, cN, below=2.2mm of s2] (s3) {3. For each module: add its controllers as an application part; call its registration extension};
\draw[ar] (s2) -- (s3);
\node[fb=5.6cm, cN, below=2.2mm of s3] (s4) {4. The host builds the application};
\draw[ar] (s3) -- (s4);
\node[fb=5.6cm, cB, below=2.2mm of s4] (s5) {5. Route-table check: no two actions share a method and a template (templates compared as written)};
\draw[ar] (s4) -- (s5);
\node[fb=5.6cm, cB, below=2.2mm of s5] (s6) {6. Scoped-filter check: each declared filter on every own-module action and on no other module's action};
\draw[ar] (s5) -- (s6);
\node[fb=5.6cm, cG, below=2.2mm of s6] (s7) {7. The host runs};
\draw[ar] (s6) -- (s7);
\node[fb=1.6cm, cR, minimum height=4.6cm, anchor=west] (stop) at ($(s2.east)!0.5!(s6.east)+(0.35,0)$) {any failed check stops start-up, in every environment};
\foreach \i in {2,5,6} \draw[ar] (s\i.east) -- (s\i.east -| stop.west);

\end{tikzpicture}
\caption{Start-up of the API host: the order of the checks, each of which stops the host in every environment.}
\label{fig:startup}
\end{figure}

Fig.~\ref{fig:startup} gives the order of the start-up checks. The module registry refuses to run on a duplicate route
prefix, controller assembly or module name, an empty or malformed name or prefix, or a module with no controllers. It
also requires each module to name a current-user middleware with a callable entry point, because a module without
one would run its requests with no tenant set, and the shared library's global tenant filter applies only when a
tenant is set, so those requests would not be scoped to a tenant; that silent defect was found by running the host (Section~\ref{sec:findings}) and turned
into a start-up error. The check is on the contract, not on the pipeline: a module that names its middleware but whose
middleware is not mounted passes it, which is fault F07 of Section~\ref{sec:rq2}. A separate assertion checks the
real route table for two actions that share a method and a template (F04). It compares templates as written, so two
templates that differ only in a parameter name are not caught. The route prefix in the module contract never
prefixes anything, and the controllers keep their own route attributes; the host uses the prefix for these checks and
to choose which requests pass through each module's exception, current-user and activity-log middleware, so a
prefix that does not match a module's real routes would give its requests the wrong identity handling.

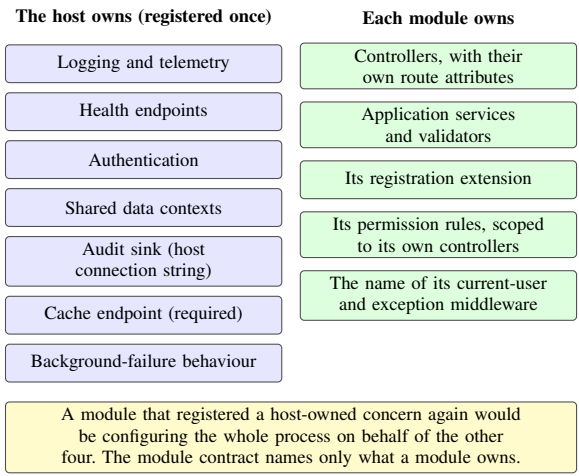
\begin{figure}[t]
\centering\begin{tikzpicture}[node distance=2mm]

\node[ttl, text width=3.5cm, align=center] (h0) at (-1.95,0) {The host owns (registered once)};
\node[ttl, text width=3.5cm, align=center] (m0) at (1.95,0) {Each module owns};
\node[fb=3.5cm, cB, below=2mm of h0] (h1) {Logging and telemetry};
\node[fb=3.5cm, cB, below=1.2mm of h1] (h2) {Health endpoints};
\node[fb=3.5cm, cB, below=1.2mm of h2] (h3) {Authentication};
\node[fb=3.5cm, cB, below=1.2mm of h3] (h4) {Shared data contexts};
\node[fb=3.5cm, cB, below=1.2mm of h4] (h5) {Audit sink (host connection string)};
\node[fb=3.5cm, cB, below=1.2mm of h5] (h6) {Cache endpoint (required)};
\node[fb=3.5cm, cB, below=1.2mm of h6] (h7) {Background-failure behaviour};
\node[fb=3.5cm, cG, below=2mm of m0] (m1) {Controllers, with their own route attributes};
\node[fb=3.5cm, cG, below=1.6mm of m1] (m2) {Application services and validators};
\node[fb=3.5cm, cG, below=1.6mm of m2] (m3) {Its registration extension};
\node[fb=3.5cm, cG, below=1.6mm of m3] (m4) {Its permission rules, scoped to its own controllers};
\node[fb=3.5cm, cG, below=1.6mm of m4] (m5) {The name of its current-user and exception middleware};
\node[fb=7.4cm, cY, anchor=north] at ($(h7.south)!0.5!(h7.south -| m5.south)+(0,-2.5mm)$) {A module that registered a host-owned concern again would be configuring the whole process on behalf of the other four. The module contract names only what a module owns.};

\end{tikzpicture}
\caption{Which concerns the host owns and registers once, and which each module owns.}
\label{fig:hostowned}
\end{figure}

\textbf{Host-owned concerns.} The host owns logging, telemetry, health, authentication and the shared data contexts,
and registers each once (Fig.~\ref{fig:hostowned}); a module that registered any of them again would be configuring
the process on behalf of the other four. Three more host-owned concerns were each found because something failed.
The audit sink, registered per module, was silently lost by whichever module omitted it, so the host now owns it and
fails at start-up without the PostgreSQL and MongoDB connection strings, the MongoDB database name, or an explicit
list of allowed cross-origin origins. The cache endpoint is required because every module calls the shared caching
registration and the permission handler resolves the cache, so an unset endpoint made every module's endpoints
answer 500 in a way that looked like an authorisation fault; the host now fails at start-up instead (F14). And because
the channels module registers an SQS queue poller, the API host is also a queue consumer; the default behaviour of the
.NET host is to stop the whole process when a background service fails, so the host sets it to ignore such failures,
and a dead poller no longer stops the API but can go unnoticed (P3). Queue polling belongs in the consumer host.

\textbf{One current-user middleware.} The services ship five current-user middleware implementations, of 48 to 117
lines, in addition to one in the shared library, and they differ in how strictly they read the tenant claim. The
edition references a single implementation, one that fails a request without a tenant, for the four modules whose
users sign in with the platform's own tokens. The channels module keeps its own, because its clients authenticate
through the external identity provider and their tenant comes from the client application, not from a claim in the
token.

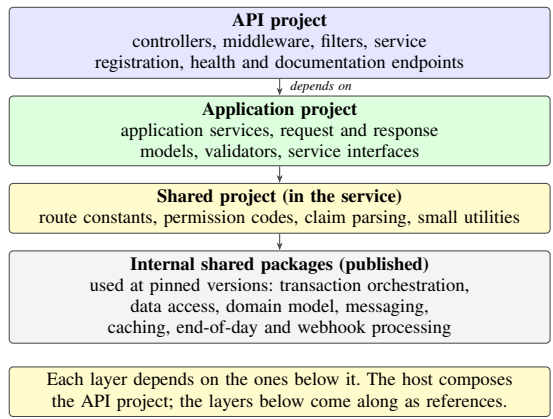
\begin{figure}[t]
\centering\begin{tikzpicture}[node distance=2mm]
\node[fb=7.0cm, cB] (s1) {\textbf{API project}\\ controllers, middleware, filters, service registration, health and documentation endpoints};
\node[fb=7.0cm, cG, below=2.2mm of s1] (s2) {\textbf{Application project}\\ application services, request and response models, validators, service interfaces};
\draw[ar] (s1) -- (s2);
\node[fb=7.0cm, cY, below=2.2mm of s2] (s3) {\textbf{Shared project (in the service)}\\ route constants, permission codes, claim parsing, small utilities};
\draw[ar] (s2) -- (s3);
\node[fb=7.0cm, cN, below=2.2mm of s3] (s4) {\textbf{Internal shared packages (published)}\\ used at pinned versions: transaction orchestration, data access, domain model, messaging, caching, end-of-day and webhook processing};
\draw[ar] (s3) -- (s4);
\node[lab, right=1mm of $(s1.south)!0.5!(s2.north)$] {depends on};
\node[fb=7.0cm, cY, below=3mm of s4] {Each layer depends on the ones below it. The host composes the API project; the layers below come along as references.};

\end{tikzpicture}
\caption{The layers inside a module, from the API project down to the internal shared packages.}
\label{fig:layers}
\end{figure}

\textbf{Middleware order.} In the transactions service the sequence is request logging, cross-origin policy, the health
endpoint, the global exception handler, routing, authentication, authorisation, the current-user middleware, an
activity-audit middleware, and then the action with its permission filter. The current-user middleware sits after
authorisation and before the audit step, so a module that omitted it would lose the user and tenant from every audit
record as well as from its queries. The host's order is path-base stripping, routing, request logging, the
cross-origin policy, authentication and authorisation, and then, on a branch selected by the module's route prefix,
the module's exception middleware, its current-user middleware and the shared activity-log middleware, before the
controllers. It preserves the relative order of the transactions service; we did not compare it with the pipelines of
the other four services, and swapping two steps of the branch was not caught by the checks (F08). Within a module the
code follows a conventional layering (Fig.~\ref{fig:layers}).

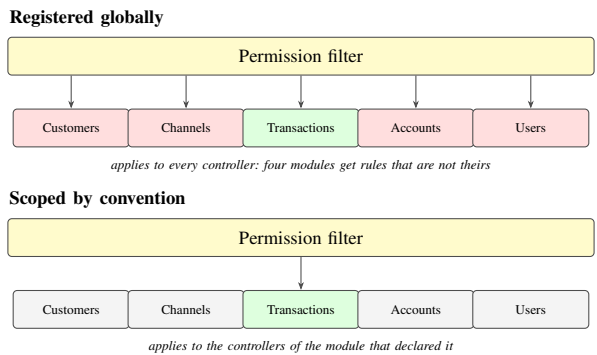
\begin{figure}[t]
\centering\begin{tikzpicture}[node distance=2mm]

\node[ttl, anchor=west] at (-3.9,0.1) {Registered globally};
\node[fb=7.6cm, cY] (g) at (0,-0.4) {Permission filter};
\foreach \m/\x/\c in {Customers/-3.04/cR, Channels/-1.52/cR, Transactions/0/cG, Accounts/1.52/cR, Users/3.04/cR} {
  \node[fb=1.38cm, \c, font=\tiny] (a\m) at (\x,-1.35) {\m};
  \draw[ar] (g.south -| a\m) -- (a\m);
}
\node[lab] at (0,-1.85) {applies to every controller: four modules get rules that are not theirs};
\node[ttl, anchor=west] at (-3.9,-2.3) {Scoped by convention};
\node[fb=7.6cm, cY] (s) at (0,-2.8) {Permission filter};
\foreach \m/\x/\c in {Customers/-3.04/cN, Channels/-1.52/cN, Transactions/0/cG, Accounts/1.52/cN, Users/3.04/cN} {
  \node[fb=1.38cm, \c, font=\tiny] (b\m) at (\x,-3.75) {\m};
}
\draw[ar] (s.south -| bTransactions) -- (bTransactions);
\node[lab] at (0,-4.25) {applies to the controllers of the module that declared it};

\end{tikzpicture}
\caption{A permission filter registered globally reaches every module; the convention restricts it to its own.}
\label{fig:scoped}
\end{figure}

\textbf{Scoped filters.} The transactions service registers its permission check as a global MVC filter, which in a
shared host would apply to every module's controllers. The collection of global filters has no notion of which module
a controller came from, so an application-model convention matches each controller's declaring assembly against the
module's controller assemblies and adds the filter to those controllers only (Fig.~\ref{fig:scoped},
Listing~\ref{lst:convention}). Its result is asserted at start-up in both directions: a filter that has leaked onto
another module's actions, and a filter missing from its own module's actions. The second cannot be seen from outside
by an unauthenticated request, which the authorisation attribute rejects before any action filter runs, which is why
it is a start-up check. In the composed host the permission filter applies to the 87 transactions actions and the
validation filter to the 369 users actions, and nothing else. The assertion covers the filters a module declares; a
module that stops declaring its filter passes (F06).

\begin{figure}[t]
\begin{lstlisting}[float=false,language={[Sharp]C},caption={The convention that scopes a module's filters to its own controllers (condensed).},label={lst:convention}]
public void Apply(ApplicationModel app)
{
    foreach (var c in app.Controllers)
    foreach (var m in _modules)
    {
        if (m.ScopedFilters.Count == 0) continue;
        var asm = c.ControllerType.Assembly;
        if (!m.ControllerAssemblies.Contains(asm)) continue;
        foreach (var f in m.ScopedFilters)
            c.Filters.Add(new TypeFilterAttribute(f));
    }
}
\end{lstlisting}
\end{figure}

\subsection{Consumer and background hosts}
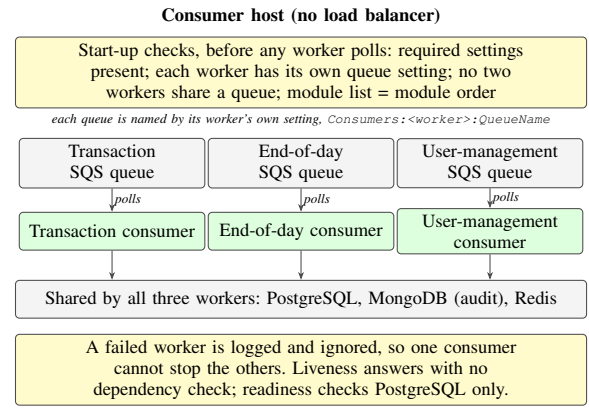
\begin{figure}[t]
\centering\begin{tikzpicture}[node distance=2mm]

\node[ttl] at (0,0.3) {Consumer host (no load balancer)};
\node[fb=7.4cm, cY] (chk) at (0,-0.45) {Start-up checks, before any worker polls: required settings present; each worker has its own queue setting; no two workers share a queue; module list = module order};
\node[lab] at (0,-1.08) {each queue is named by its worker's own setting, \texttt{Consumers:<worker>:QueueName}};
\foreach \n/\t/\x in {t/Transaction/-2.5, e/End-of-day/0, u/User-management/2.5} {
  \node[fb=2.3cm, cN] (q\n) at (\x,-1.65) {\t{} SQS queue};
  \node[fb=2.3cm, cG] (w\n) at (\x,-2.55) {\t{} consumer};
  \draw[ar] (q\n) -- node[lab, right] {polls} (w\n);
}
\node[fb=7.4cm, cN] (sh) at (0,-3.45) {Shared by all three workers: PostgreSQL, MongoDB (audit), Redis};
\foreach \n in {t,e,u} \draw[ar] (w\n) -- (w\n |- sh.north);
\node[fb=7.4cm, cY, below=2mm of sh] {A failed worker is logged and ignored, so one consumer cannot stop the others. Liveness answers with no dependency check; readiness checks PostgreSQL only.};

\end{tikzpicture}
\caption{The consumer host: three workers, each on its own queue, behind start-up checks.}
\label{fig:consumer}
\end{figure}

The consumer host composes three workers through a smaller contract of its own: each worker names itself, the
configuration key that holds its queue name, and a call that registers it (Fig.~\ref{fig:consumer}). The background
host composes a single service and uses no contract; a contract exists to stop several services configuring one
process on each other's behalf. The hosts are separate processes on purpose: a long end-of-day run must not be killed
by an API deployment, and an API restart must not interrupt message processing.

The consumer host's start-up checks were each added after a failure that gave no error. Every worker used to read one
process-wide queue-name setting, which is right while a service owns its process and wrong when three share one: all
three would poll the same queue and each would pull and drop the others' messages, with nothing logged. The host
therefore requires each worker to have its own queue setting and refuses to start if one is unset (F13) or two
workers point at the same queue (F12). It asserts that the list of workers and the module order that decides
configuration precedence agree, and it requires the database, document-store and cache settings before any worker
starts, because a consumer has no request to fail. The document-store database name has two keys and the deployed
secret carries only one, so the host accepts either. Ignoring a failed background service suits this host: one
failing consumer should not stop the other two.

Composing the workers forced three changes. Two model assemblies of one name cannot share a process, so one
service's model assembly was renamed, in its assembly name only. The whole consumer host runs with preview language
features enabled, because one worker's project requires them and the setting propagates to every caller. And document
generation in the users worker does not work in this host, because its standalone start-up awaits a browser download
before building the host; replicating that would hold the other two workers behind a download at every start (X5).
Placement can also follow from a process-wide registration: the users worker registers the identity framework, which
makes its cookie schemes the default authentication scheme for the whole process. In the API host that would compete
with the bearer-token default every module relies on; the consumer host has no authentication pipeline to disturb.

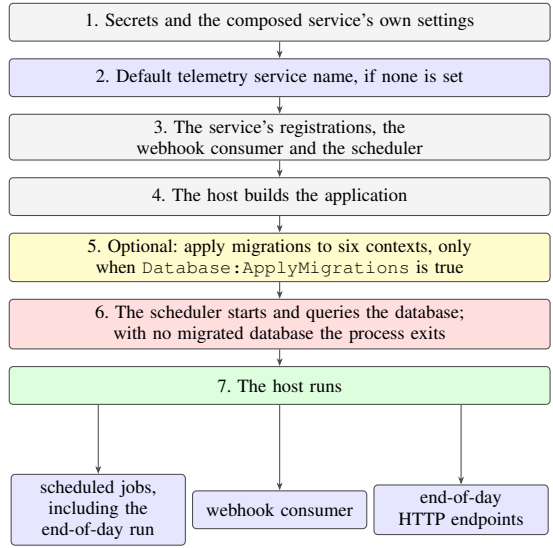
\begin{figure}[t]
\centering\begin{tikzpicture}[node distance=2mm]
\node[fb=7.0cm, cN] (s1) {1. Secrets and the composed service's own settings};
\node[fb=7.0cm, cB, below=2.2mm of s1] (s2) {2. Default telemetry service name, if none is set};
\draw[ar] (s1) -- (s2);
\node[fb=7.0cm, cN, below=2.2mm of s2] (s3) {3. The service's registrations, the webhook consumer and the scheduler};
\draw[ar] (s2) -- (s3);
\node[fb=7.0cm, cN, below=2.2mm of s3] (s4) {4. The host builds the application};
\draw[ar] (s3) -- (s4);
\node[fb=7.0cm, cY, below=2.2mm of s4] (s5) {5. Optional: apply migrations to six contexts, only when \texttt{Database:ApplyMigrations} is true};
\draw[ar] (s4) -- (s5);
\node[fb=7.0cm, cR, below=2.2mm of s5] (s6) {6. The scheduler starts and queries the database; with no migrated database the process exits};
\draw[ar] (s5) -- (s6);
\node[fb=7.0cm, cG, below=2.2mm of s6] (s7) {7. The host runs};
\draw[ar] (s6) -- (s7);
\foreach \t/\x [count=\i] in {{scheduled jobs, including the end-of-day run}/-2.4, {webhook consumer}/0, {end-of-day HTTP endpoints}/2.4} {
  \node[fb=2.15cm, cB] (o\i) at (\x,-6.45) {\t};
  \draw[ar] (s7.south -| o\i) -- (o\i);
}

\end{tikzpicture}
\caption{Start-up of the background host, and what it runs.}
\label{fig:background}
\end{figure}

The background host mirrors its standalone service closely (Fig.~\ref{fig:background}), and its few differences were
found by running it. The service's telemetry set-up passes the service-name environment variable to a call that throws
when it is empty; the standalone service gets the variable from a launch profile, so the failure appears only where
nothing sets it, such as a task definition, as a crash loop whose message names telemetry and not the missing
variable. The host supplies a default before the service's registrations run. It exposes the same liveness and
readiness paths as the other hosts, and it does not redirect to HTTPS, because TLS ends at the load balancer and a
redirect inside the container would send the health probe to a port nothing listens on. The end-of-day batch shows
why the split is needed: it can be started from an HTTP endpoint, but its jobs are dispatched and consumed through
SNS topics and SQS queues by a hosted service, so the long-running work belongs to the consumer and background hosts.

\subsection{Start-up behaviour}
Start-up with the database configured but unreachable was observed for the earlier report (Table~\ref{tab:startdb}).
The API and consumer hosts start, answer liveness with 200 and report not ready with 503: a data context opens its
connection only when first used, so registering the contexts does not need the database, while the readiness probe
queries it. The background host does not start: its scheduler seeds its cron definitions during start-up and
queries the database, so it exits and, in a container service, crash-loops until the database is reachable and
migrated. A database that is not configured stops all three. The earlier report stated that readiness also covers
MongoDB and Redis; in the code studied here it covers PostgreSQL only (X3), so an unreachable audit store or cache
does not take a host out of rotation, and Section~\ref{sec:discussion} proposes a readiness contract.

\begin{table}[t]
\centering\footnotesize
\caption{Start-up with the database configured but unreachable, as observed for the earlier report (E6).}
\label{tab:startdb}
\begin{tabular}{@{}lccc@{}}
\toprule
Host & Starts & Liveness & Readiness\\
\midrule
API & yes & 200 & 503\\
Consumer & yes & 200 & 503\\
Background & no: it exits & -- & --\\
\bottomrule
\end{tabular}
\end{table}

% =====================================================================================================
\section{The Shared Package and the Dependency Policy}
\label{sec:deps}

The shared library is consumed exactly as the services consume it: as published packages from a private feed, at the
version pinned in the central package file. A script asserts that all 195 pins of the shared packages across the eight
repositories agree (it caught F01). This matters even with a published package, because one process loads one
assembly version, so a single stale pin is a run-time type-load failure and not a build error. Before the shared
package was published, the edition used a locally packed prerelease of version 1.1.0 while upstream published on the
1.0 line; semantic versioning sorts a prerelease below the release of the same number, so a pin that lost its suffix
resolved the public package and the build succeeded with none of the shared fixes in it. Nothing errored; the
tenant-aware business date was simply absent.

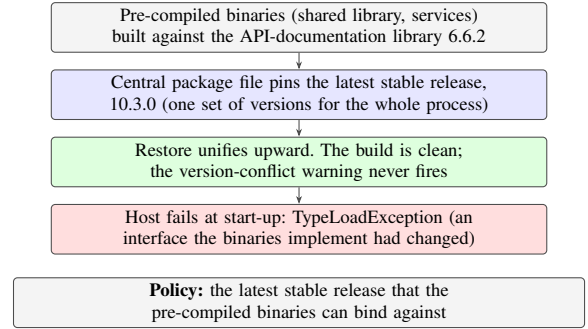
\begin{figure}[t]
\centering\begin{tikzpicture}[node distance=2mm]
\node[fb=6.4cm, cN] (s1) {Pre-compiled binaries (shared library, services) built against the API-documentation library 6.6.2};
\node[fb=6.4cm, cB, below=2.2mm of s1] (s2) {Central package file pins the latest stable release, 10.3.0 (one set of versions for the whole process)};
\draw[ar] (s1) -- (s2);
\node[fb=6.4cm, cG, below=2.2mm of s2] (s3) {Restore unifies upward. The build is clean; the version-conflict warning never fires};
\draw[ar] (s2) -- (s3);
\node[fb=6.4cm, cR, below=2.2mm of s3] (s4) {Host fails at start-up: TypeLoadException (an interface the binaries implement had changed)};
\draw[ar] (s3) -- (s4);
\node[fb=7.4cm, cN, below=3mm of s4] {\textbf{Policy:} the latest stable release that the pre-compiled binaries can bind against};

\end{tikzpicture}
\caption{How a version pin above the binding ceiling passes the build and fails at start-up (a recorded example).}
\label{fig:ceiling}
\end{figure}

Central package management governs the edition's repository, while the referenced service projects keep their own
older pins, and the restore graph is unified upward to the edition's versions. Module code compiled against an older
major version then runs against a newer one, which fails as a type-load or missing-method error on a code path, not
as a failed build. The edition makes the one compile-time shadow of this, the warning about conflicting versions of a
dependent assembly, an error. The policy is therefore the latest stable release that the pre-compiled binaries can
bind against. This was learned the hard way (Fig.~\ref{fig:ceiling}): pinning the then-latest major version of the API
documentation library made the host fail at start-up, because an interface implemented in the shared library had
changed, while the build was clean and the version-conflict warning never fired. A sweep then found ten major-version
jumps above that ceiling; eight are pinned down, and the other two are test-only packages not yet used.

Two known advisories remained in the composed image at the time of the earlier report (E6). One, rated high, affects
the object-mapping library at the version in use; there is no patched release in that major version, and the next lies
above the binding ceiling, because two modules were compiled against one major version and a third against an
earlier one, so it is recorded as a decision and not an oversight. The other, rated moderate, affects a deprecated
tracing exporter that arrives transitively and has no fix. Five other advisories were resolved; four arrived
transitively, so entries in the central package file did nothing until the host referenced the packages explicitly.

% =====================================================================================================
\section{Configuration}
\label{sec:config}

Each composed service keeps its configuration in its own repository, and the hosts load those files instead of
restating them. The build links each service's settings files into a per-module folder in the output, and the hosts
apply them in a declared order. Precedence runs from the modules in the declared order, through the host's own
settings files, to user secrets, environment variables and the command line, so a later module overrides an earlier
one and the host always has the last word (Fig.~\ref{fig:precedence}). Mechanically the host inserts each module's
files ahead of its own (Listing~\ref{lst:config}, Fig.~\ref{fig:configbuild}), reading them from the output directory,
because the content root of a development run is the project directory, where they do not exist. The declared module
order is asserted at start-up to match the module list (F03).

\begin{figure}[t]
\centering\begin{tikzpicture}[node distance=2mm]

\foreach \t/\c [count=\i] in {{Modules, in the host's declared order (a later module overrides an earlier one)}/cN,
                              {Host settings files: appsettings.json, appsettings.\{Environment\}.json}/cG,
                              User secrets/cB, Environment variables/cB, Command line/cB} {
  \node[fb=6.4cm, \c, minimum height=0.55cm] (p\i) at (0.4,0.68*\i) {\t};
}
\draw[ar] (-3.1,0.3) -- node[lab, sloped, above] {higher precedence} (-3.1,3.7);
\node[lab, align=center] at (0.2,-0.05) {each host logs the keys that module order had to decide};

\end{tikzpicture}
\caption{Configuration precedence, from lowest to highest.}
\label{fig:precedence}
\end{figure}
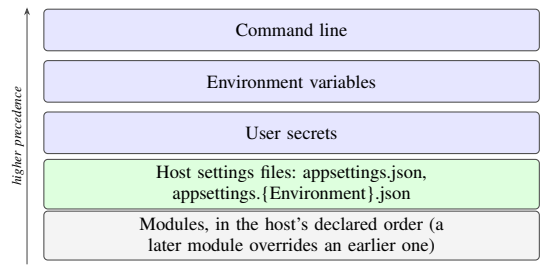

\begin{figure}[t]
\centering\begin{tikzpicture}[node distance=2mm]
\node[fb=7.0cm, cN] (s1) {1. The secrets source is loaded first, before anything reads configuration};
\node[fb=7.0cm, cN, below=2.2mm of s1] (s2) {2. The host declares the module order: Customers, Channels, Transactions, Accounts, Users};
\draw[ar] (s1) -- (s2);
\node[fb=7.0cm, cB, below=2.2mm of s2] (s3) {3. For each module, in that order, its settings files (\texttt{modules/<Module>/appsettings.json} and the environment file) are inserted ahead of the host's own, read from the output folder, not reloaded on change};
\draw[ar] (s2) -- (s3);
\node[fb=7.0cm, cG, below=2.2mm of s3] (s4) {4. The host's own settings files, user secrets, environment variables and the command line are applied above them};
\draw[ar] (s3) -- (s4);
\node[fb=7.0cm, cR, below=2.2mm of s4] (s5) {5. Check: the module list equals the declared order, or the host refuses to start};
\draw[ar] (s4) -- (s5);
\node[fb=7.0cm, cY, below=2.2mm of s5] (s6) {6. After the build, the host logs each file loaded and every key that two modules set differently, with the winner};
\draw[ar] (s5) -- (s6);
\end{tikzpicture}
\caption{How the host builds its configuration at start-up, in order.}
\label{fig:configbuild}
\end{figure}
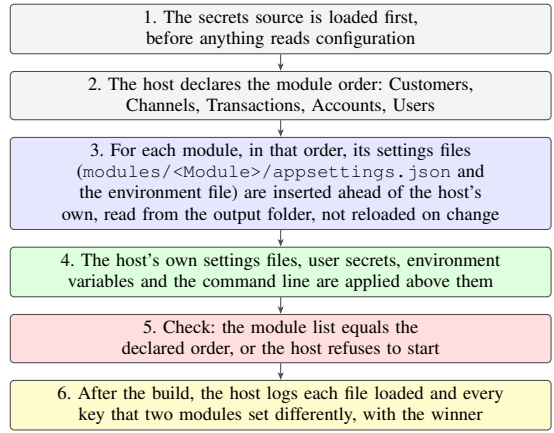

\begin{figure}[t]
\begin{lstlisting}[float=false,language={[Sharp]C},caption={Inserting each module's settings files below the host's (condensed).},label={lst:config}]
// Each module's files, in the order the host declares the
// modules, go in AHEAD of the host's own settings files.
var index = IndexOfHostAppSettings(config.Sources);
foreach (var module in moduleNames)
foreach (var name in new[] {
    "appsettings.json", $"appsettings.{env}.json" })
{
    var path = $"modules/{module}/{name}";
    if (!File.Exists(Path.Combine(root, path))) continue;
    config.Sources.Insert(index++,
        new JsonConfigurationSource {
            Path = path, Optional = true,
            ReloadOnChange = false, FileProvider = provider });
}
\end{lstlisting}
\end{figure}

\begin{figure}[t]
\centering\begin{tikzpicture}[node distance=2mm]

\node[ttl] (t1) at (0,0.15) {Before: every service's environment file targeted one path};
\foreach \m/\x in {Customers/-3.0, Channels/-1.5, Transactions/0, Accounts/1.5, Users/3.0} {
  \node[fb=1.38cm, cN, font=\tiny] (e\m) at (\x,-0.45) {\m\\ env file};
}
\node[fb=5.6cm, cR, below=6mm of eTransactions] (one) {one output path: \texttt{appsettings.<Environment>.json}};
\foreach \m in {Customers, Channels, Transactions, Accounts, Users} \draw[ar] (e\m.south) -- (one.north);
\node[fb=7.4cm, cR, below=2mm of one] (lost) {Only one file survives, chosen by reference order: the users service's, with its mail key, webhook addresses, cache password and token-signing secret, copied next to the host's binaries.};
\draw[ar] (one) -- (lost);
\draw[dashed, black!50] ($(lost.south west)+(0,-2mm)$) -- ($(lost.south east)+(0,-2mm)$);
\node[ttl, below=4mm of lost] (t2) {After: each service's files get a folder of their own};
\foreach \m/\x in {Customers/-3.0, Channels/-1.5, Transactions/0, Accounts/1.5, Users/3.0} {
  \path (t2.south) ++(0,-2mm) coordinate (row);
  \node[fb=1.38cm, cG, font=\tiny, anchor=north] (f\m) at (\x,0 |- row) {\texttt{modules/}\\ \m/};
}
\node[fb=7.4cm, cY, below=3mm of fTransactions] (ok) {The host applies every module's files on purpose, in the declared order, below its own settings. Nothing is overwritten, and the result no longer depends on build order.};
\draw[ar] (fTransactions) -- (ok);

\end{tikzpicture}
\caption{The settings-file collision before the edition applied module files deliberately, and what replaced it.}
\label{fig:collision}
\end{figure}
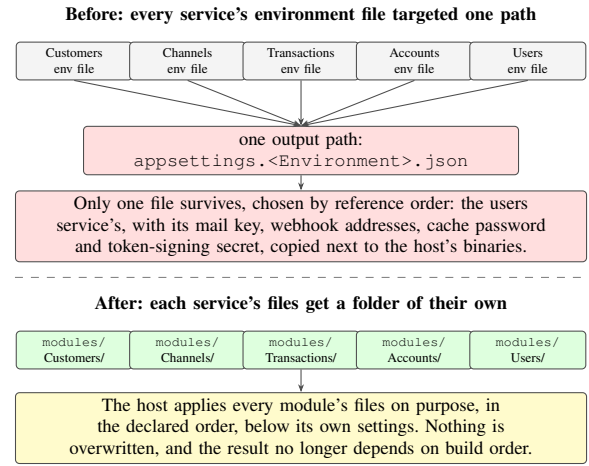

This replaced behaviour that was accidental (Fig.~\ref{fig:collision}). All five services ship an
environment-specific settings file as content; the build flows a referenced project's content into the referencing
project, and every file targeted the same output path, so one service's environment file silently became the host's,
decided by reference order. It was the users service's file, which also meant its mail-provider key, chat webhook
addresses, cache password and token-signing secret were copied next to the host's binaries.

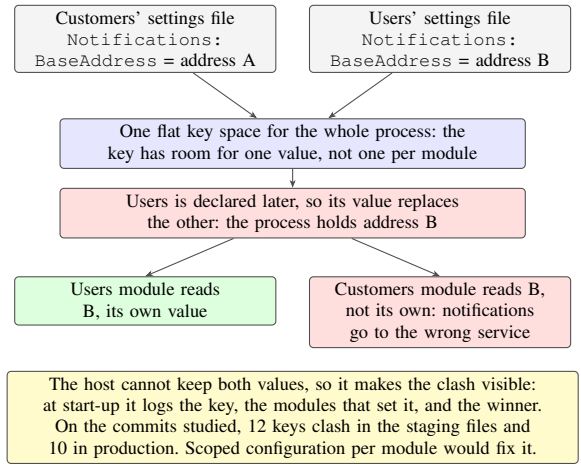
\begin{figure}[t]
\centering\begin{tikzpicture}[node distance=2mm]

\node[fb=3.3cm, cN] (a) at (-1.95,0) {Customers' settings file\\ \texttt{\scriptsize Notifications:}\\ \texttt{\scriptsize BaseAddress} = address A};
\node[fb=3.3cm, cN] (b) at (1.95,0) {Users' settings file\\ \texttt{\scriptsize Notifications:}\\ \texttt{\scriptsize BaseAddress} = address B};
\node[fb=6.0cm, cB, below=6mm of {$(a.south)!0.5!(b.south)$}] (k) {One flat key space for the whole process: the key has room for one value, not one per module};
\draw[ar] (a) -- (k); \draw[ar] (b) -- (k);
\node[fb=6.0cm, cR, below=2.5mm of k] (w) {Users is declared later, so its value replaces the other: the process holds address B};
\draw[ar] (k) -- (w);
\node[fb=3.3cm, cG, anchor=north] (u) at ($(w.south)+(-1.95,-5mm)$) {Users module reads B, its own value};
\node[fb=3.3cm, cR, anchor=north] (c) at ($(w.south)+(1.95,-5mm)$) {Customers module reads B, not its own: notifications go to the wrong service};
\draw[ar] (w) -- (u.north); \draw[ar] (w) -- (c.north);
\node[fb=7.4cm, cY, anchor=north] (n) at ($(c.south -| w.south)+(0,-3mm)$) {The host cannot keep both values, so it makes the clash visible: at start-up it logs the key, the modules that set it, and the winner. On the commits studied, 12 keys clash in the staging files and 10 in production. Scoped configuration per module would fix it.};

\end{tikzpicture}
\caption{Why a flat key space cannot serve two modules that want different values for one key.}
\label{fig:flatkey}
\end{figure}

Layering is not a merge. All modules share one flat key space, so when two modules set the same key one value
replaces the other, and a module cannot be given a value of its own (Fig.~\ref{fig:flatkey}). Each host logs every key
that module order had to decide. On the commits studied here the modules' staging files disagree on 12 keys and their
production files on 10 (the earlier report counted nine): the two end-of-day topic settings, which different modules
point at different cloud accounts; the notification base address and path; the sender address and name; an S3 bucket;
a transfer VAT setting; and the default end dates of two features. These are not formatting differences: with this
ordering, two modules' notifications go to a different service at a different path and their end-of-day messages to
a different cloud account. Scoped configuration per module would fix it; until then the start-up warnings make the
resolution visible.

% =====================================================================================================
\section{The Database}
\label{sec:database}

The edition requires no schema change, by construction. Its repository defines no data context and no migration, and
no type in it derives from a data context. The five shared contexts (transaction, accountmaster, customer, user and
card schemas, each with its own migrations history table) come from the shared packages at the version the services
use, and all five run on one connection per request. The modules reference the service projects as they are, so no
entity, configuration or mapping changes, both editions read and write the same schema, and either can be pointed at
the other's database; this study ran both on one schema and compared their ledgers. By default no host applies
migrations at run time; the background host applies every migration at start-up when a flag is set.

A fresh database takes three steps with different owners (Fig.~\ref{fig:bootstrap}, Listing~\ref{lst:bootstrap}).
The flag makes the background host migrate six contexts, the five shared ones and the scheduler's, because it already
registers all six, cannot start against an unmigrated database, and is naturally single-instance. Reference data that a
request cannot proceed without (permissions, the super-administrator role, an active user, currencies, countries, the
head-office branch) is in no migration; it is seeded by the users service's initialiser, which the API host runs once
when a flag is set, with passwords supplied through configuration. The scheduler's cron rows must be created once by
hand. The emulation in this study used exactly this path to build its seed snapshot.

\begin{figure}[t]
\centering\begin{tikzpicture}[node distance=2mm]
\node[fb=7.2cm, cB] (s1) {\textbf{1. Schema.} Background host with \texttt{Database:ApplyMigrations=true} applies all six contexts: the five shared contexts and the scheduler's};
\node[fb=7.2cm, cG, below=2.2mm of s1] (s2) {\textbf{2. Reference data.} API host run once with \texttt{Database:SeedData=true} (users initialiser): permissions, super-administrator role, active user, currencies, countries, enumerations, bank parameters, head-office branch, customer-number sequences};
\draw[ar] (s1) -- (s2);
\node[fb=7.2cm, cN, below=2.2mm of s2] (s3) {\textbf{3. Schedules.} Create the scheduler's cron rows once (dashboard or API); \texttt{InitiateEod} in particular, or no end-of-day run starts};
\draw[ar] (s2) -- (s3);
\node[fb=7.2cm, cR, below=2.2mm of s3] (s4) {\textbf{Still manual:} a client-application row matching the token's \texttt{client\_id}, and a user named CHANNELS for the tenant (channels module)};
\draw[ar] (s3) -- (s4);
\end{tikzpicture}
\caption{Bootstrapping a fresh environment: three steps with different owners, and what remains manual.}
\label{fig:bootstrap}
\end{figure}
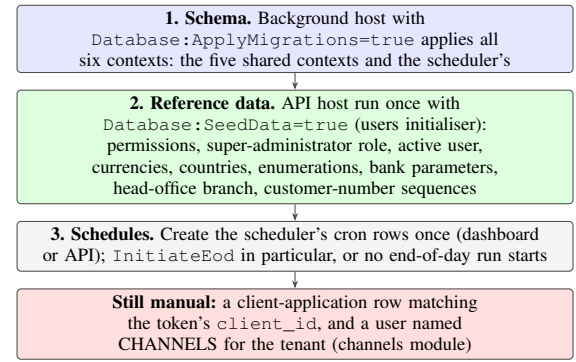

\begin{figure}[t]
\begin{lstlisting}[float=false,caption={Bootstrapping a fresh database in three steps (condensed).},label={lst:bootstrap}]
# 1. Schema: every migration, all six contexts
#    (background host).
Database__ApplyMigrations=true dotnet run --project \
  src/Host/CoreBanking.Platform.BackgroundServices
# 2. Reference data: run the users API once, flag on.
cd ../CoreBanking.UserManagement
Database__SeedData=true Seed__AdminPassword=... \
  Seed__ChannelsPassword=... \
  dotnet run --project src/UserManagement/Api
# 3. Schedules: create the scheduler's cron rows once
#    (dashboard or API); InitiateEod in particular,
#    or no end-of-day run starts.
\end{lstlisting}
\end{figure}

Two problems appeared when migrations were applied. The scheduler's data context has model changes that its own
committed migration does not contain, so the migration tooling refuses to run; the host suppresses that one warning
for that context only, which applies the committed migrations as an earlier scripted workaround did but does not
remove the cause, which needs a new migration in the scheduler's repository. And an index drop in a migration is
unconditional while index names drift between environments: one environment had a singular name where the
migrations drop a plural one, and three such names across 162 drop targets stopped a run twice. The drift script
(Section~\ref{sec:checks}) is run before the flag is enabled against an existing database.

% =====================================================================================================
\section{Study Design}
\label{sec:design}

This is a single-case experience report \cite{runeson2009guidelines} with embedded controlled experiments. The
unit of analysis is the edition as a replacement deployment for the nine-service edition. We ask:
\textbf{RQ1}: within the tested coverage, does the composed edition behave like the nine-service edition for reads,
financial writes and access control? \textbf{RQ2}: which of a set of injected composition faults does the edition's
own check suite detect, and which does it miss? \textbf{RQ3}: what does composition change operationally, in the
emulation: behaviour under failure, performance under the real task sizes, and estimated cost?

\subsection{Emulated AWS deployment}
The emulation mirrors the production deployment in what the services and hosts themselves see: the same binaries,
built from the same commits; the same task sizes, with CPU and memory limits enforced per task; the same gateway and
load-balancer path rules; the same topic-to-queue subscriptions; and configuration built from the services' own
settings. It does not reproduce the network between tasks, the latency and throttling of the managed services, or
the isolation of separate machines (Section~\ref{sec:threats}).
Both editions were built and published on .NET 10 (SDK 10.0.301, runtime 10.0.9) from the same commits of the eight
repositories, on copies of the repositories, and run from their publish output (what the images contain).
Table~\ref{tab:emulation} lists what stands in for each AWS component. Every task is one process whose Fargate CPU
share is enforced as Fargate's cgroup enforces it, by CFS bandwidth control: a controller reads the process's
accumulated CPU time from the kernel every 2\,ms and stops the process once it has used vCPU $\times$ 100\,ms in the
current 100\,ms period, carrying any overshoot into the next. Under saturating load it held the API host at
0.96--0.97 cores against its 1-vCPU quota; a duty-cycle limiter we tried first (\code{cpulimit}) let the same
process use 1.62 cores, and every timing measured with it was discarded and re-run. Each task also has
\code{DOTNET\_PROCESSOR\_COUNT=1}, as .NET reports inside a container of at most one vCPU, and
\code{GCHeapHardLimit} at 75\% of task memory, the .NET default in a memory-limited container. The real gateway runs
in front with its routes pointed at the emulated ALB. Outbound traffic from the tasks passes through an egress guard
that refuses anything not local, so no real infrastructure was contacted. Everything ran on a single host with 10 CPU
cores and 16\,GiB of memory; tool versions are listed in Table~\ref{tab:versions}.

\begin{table}[t]
\centering\footnotesize
\caption{Emulation of the AWS deployment.}
\label{tab:emulation}
\begin{tabularx}{\columnwidth}{@{}lL@{}}
\toprule
AWS component & Emulation\\
\midrule
ECS Fargate task & process from publish output; CFS-style CPU quota at the task's vCPU per 100\,ms; heap limit 75\% of task memory\\
Internal ALB & nginx 1.31 with the same path rules; default 10\,s fail timeout per target\\
API gateway & the real Ocelot gateway, routes rewritten to the emulated ALB\\
SNS, SQS, Secrets Manager, S3 & moto 5.2 in server mode \cite{moto}, same topic-to-queue subscriptions\\
Managed data stores & PostgreSQL 16.15, MongoDB 8.0, Redis 8.10 on localhost\\
Configuration & one secret per edition built from the services' own settings, local endpoints substituted\\
\bottomrule
\end{tabularx}
\end{table}

Every experiment starts from one database snapshot, migrated and seeded by the edition's own bootstrap path, with
empty caches, queues and audit store. Tokens are minted with the users service's signing scheme, with claims for
users in the seeded database.

\subsection{Instruments}
\textbf{Workload.} A script runs 47 named cases through the gateway. Twenty-one build a minimal bank, each step made
by one administrator and approved by another: transaction limits for five types, role and user limit grants, three
general-ledger accounts, three internal products and six internal accounts. Twenty-six exercise the posting
engine: ten balanced postings, a multi-leg posting, five invalid postings, a re-post of a posted identifier, the same
posting sent twice without an identifier, first and second reversals, a multi-leg reversal, the maker-checker save, authorise,
re-authorise and self-authorise path, and 20 concurrent debits of one account. Each case has an expected outcome
written before the runs; one (the posting sent twice) records behaviour without asserting it, and one (B25) was
corrected after the runs, when the platform owner confirmed that a user holding both the save and the authorise
permission may authorise their own transaction. The observed behaviour did not change; the raw results keep the
original verdict, and the scoring script applies the corrected one. The appendix lists all 47.

\textbf{Ledger check.} After each run the legs and balances of the six accounts are read from PostgreSQL and checked
with exact decimals: debits equal credits per transaction, legs are numbered without gaps, each leg moves its
balance by exactly its amount, each account's legs form one chain from zero (linked by balance, since under
concurrency leg identifiers are assigned before the account lock is taken), each clear balance equals the end of
its chain, and reversals net to zero. Transactions with identical legs are listed for review as duplicate candidates, not reported as findings, since
identical postings under different transaction identifiers are legitimate.

\textbf{Read probes.} Every parameterless GET in the OpenAPI document (231 paths) is called through the gateway by
five principals: a super-administrator; no token; a valid token for a user with no role; a super-administrator
token without a tenant claim; and one with an unknown tenant.

\textbf{Normalisation.} Two independent runs cannot produce byte-identical responses, because the platform
generates identifiers and timestamps. Before comparison, response bodies are serialised with sorted keys and four
classes of value are replaced by a placeholder: UUIDs, ISO-8601 timestamps, W3C trace identifiers, and 32-digit
hexadecimal concurrency stamps, plus the fields \code{traceId}, \code{requestId}, \code{timestamp},
\code{correlationId} and \code{jti}. Amounts, balances, counts, names, statuses, error messages and the structure
of every body are compared as they are. Ledger legs are compared after dropping only database identifiers and
replacing account numbers by account names; every amount and every balance before and after each leg is compared
exactly. Two checks bound what normalisation can hide. In fresh runs with injected faults that turned out not to
change observable behaviour (F10, F11 in Section~\ref{sec:rq2}), the comparison reported no difference on any of the
1,155 probes, so normalisation removes run-to-run noise without leaving false alarms. In three fault runs (F07, F08,
F16) it reported 13 differences in bodies whose status code was unchanged, so it keeps differences of content.

\textbf{Failure, fault and load harnesses} are described with their results.

\textbf{Current-user middleware in the measurements.} The measurements of Sections~\ref{sec:rq1}
to~\ref{sec:rq3perf} were taken with each module's original current-user middleware. The workload, the ledger check and
the read probes were then repeated with the single middleware of Section~\ref{sec:composition} (78 legs, debits equal
to credits at 243,248.08, as before); Section~\ref{sec:rq1} separates the original defect, the change and these
tests.

\subsection{Evidence categories}
Table~\ref{tab:evidence} states what kind of evidence supports each claim.

\begin{table}[t]
\centering\footnotesize
\caption{Evidence categories used in this paper.}
\label{tab:evidence}
\begin{tabularx}{\columnwidth}{@{}clL@{}}
\toprule
& Category & Used for\\
\midrule
E1 & Source inspection & composition method, configuration, task definitions\\
E2 & Automated check, repository script & route parity, pins, corrections, start-up assertions\\
E3 & Controlled experiment, emulated AWS & equivalence, failures, fault injection, performance\\
E4 & Price model from AWS list prices & cost (an estimate, not a bill)\\
E5 & Projection to AWS from configuration & recovery times on a real ALB (not measured)\\
E6 & Earlier report, deployed environment & observations recorded for the earlier version; one ledger export\\
\bottomrule
\end{tabularx}
\end{table}

% =====================================================================================================
\section{RQ1: Tested Behavioural Equivalence}
\label{sec:rq1}

\textbf{Route parity (E2).} The route-parity script diffs each standalone service's own OpenAPI document with the
composed one under that module's prefix, by method and path. All 844 operations are present in the composed API with
none missing or extra: users 369, accounts 288, transactions 87, channels 67, customers 33. (The earlier report gave
838, with accounts at 282; the accounts service has moved since, and its new operations reached both editions with
no edit in the edition's repository.) Route parity says that the operations exist, not that they behave alike.

\textbf{Financial writes (E3).} Table~\ref{tab:workload} keeps four measures apart, because they answer different
questions. \emph{Equivalence} asks whether the two editions behaved alike; \emph{ledger consistency}, whether each
ledger obeys double-entry rules; \emph{harness assertions}, whether each response met the expectation the workload
script encoded; and \emph{business-rule compliance}, whether the behaviour met the intended financial and authorisation
rules, judged independently of the script. The last two are kept apart because a harness can only check what its
author thought to encode; with the corrected expectation for B25 they agree in every case. The workload ran six times per edition (the main run and the workload phase of each of the five failure
repetitions), each on a fresh deployment, with the same counts every time. \emph{Equivalence}: all 47
cases returned the same status in both editions in all six runs, and the normalised ledgers, including every balance before and
after every leg, and the final balances of all six accounts, are identical. \emph{Consistency}: both ledgers pass
every check with no finding; twenty concurrent debits of one account all succeeded and moved its balance by exactly
$-2{,}000.00$ along an unbroken chain, so the Redis lock serialised them in both editions. \emph{Harness assertions}: 47 of 47 met in both editions. \emph{Business-rule compliance}: 47 of 47 in both editions. In B25 a user who holds both the
permission to save a transaction and the permission to authorise it authorised one they had saved; the platform
intends this.
Segregation of duties is enforced through roles: authorising requires the \code{AUTHORIZE\_TXN} permission, a user
without it is refused, and a bank that wants every transaction approved by a second person assigns the save and
authorise permissions to different roles.
An already posted transaction cannot be posted again (B17): the attempt was refused and nothing was posted. A
posted transaction can only be reversed, which the workload exercised for a two-leg and a four-leg transaction
(B19, B21); a second reversal of the same transaction was refused (B20). Sending the same posting twice without an identifier produced two
transactions with different transaction identifiers (B18), as intended: two identical postings are two genuine
transactions. Postings are
made for new transactions or for transactions in the Entered state; a posted transaction is never posted again.

\begin{table}[t]
\centering\footnotesize
\caption{Workload results, as four separate measures; identical in each of six runs per edition.}
\label{tab:workload}
\begin{tabularx}{\columnwidth}{@{}Lr@{}}
\toprule
Measure & Result\\
\midrule
\multicolumn{2}{@{}l}{\emph{Equivalence between the editions}}\\
Cases with the same status (each of 6 runs) & 47 of 47\\
Normalised legs, balances and final balances identical & yes\\
Repeated with the single current-user middleware & identical\\
\multicolumn{2}{@{}l}{\emph{Consistency of each ledger (each edition)}}\\
Posted legs; transactions & 78; 37\\
Total debits = total credits & 243,248.08\\
Ledger-check findings (balance, chain, reversal, numbering) & 0\\
Hot account after 20 concurrent debits of 100.00 & $-2{,}000.00$\\
\multicolumn{2}{@{}l}{\emph{Harness assertions (each edition)}}\\
Cases meeting the expectation the script encoded & 47 of 47\\
\multicolumn{2}{@{}l}{\emph{Business-rule compliance (each edition)}}\\
Cases complying with the intended rules & 47 of 47\\
Invalid postings rejected & 5 of 5\\
Second reversal rejected & yes (422)\\
User with save and authorise permissions authorises own transaction & allowed (by design)\\
Same posting sent twice, no identifier & two transactions, distinct identifiers (by design)\\
Re-post of a posted transaction & refused; reversal is the only route\\
\bottomrule
\end{tabularx}
\end{table}

\textbf{Reads and access control (E3).} Each edition answered 1,155 probe requests. For every principal, 230 of the
231 paths returned the same status in both editions, and every 200 response had an identical normalised body (114
for the administrator). The one difference, \code{GET /api/v1/accounts}, comes from the emulated load balancer, not from either edition.
In the nine-service edition this bare path reaches the nginx rule for the accounts service, which is written for
the prefix with a trailing slash, so nginx answers with a redirect; following it loses the token, and every principal
receives 401. An ALB path rule forwards a matching request without redirecting, and the accounts service called
directly, without the load balancer, answers the administrator with the same 404 as the composed edition. The
difference therefore says nothing about the deployment on a real ALB, which this study did not measure; we report
it rather than drop the path. Apart from that one path, the two editions gave the same response to every principal
on every probed path, and without a token 230 paths return 401 in both.

\textbf{Tenant context: defect, change and tests (E2, E3).} \emph{Defect.} When the API host was first run, no
current-user middleware was registered, so no tenant was set; the shared global tenant filter applies only when a
tenant is set, so requests ran without tenant scoping and still answered 200 (Section~\ref{sec:findings}). The
services' own middleware implementations also differ in how strictly they read the tenant claim
(Section~\ref{sec:composition}). \emph{Change.} The host now mounts a current-user middleware for every module, the
module registry refuses to start without one, and the four token-based modules use one implementation that refuses
a request without a tenant claim; the channels module keeps its own. \emph{Tests after the change.} The start-up
check was exercised by fault injection (F07 shows its limit: a middleware that is named but not mounted passes it).
The workload, the ledger check and the read probes were repeated with the single middleware: all 47 expectations
were met with identical statuses, the ledger was identical leg for leg, the responses to the other four principals
were unchanged, and a request without a tenant was refused before it reached any data on all 231 paths. The other
measurements of Sections~\ref{sec:rq1} to~\ref{sec:rq3perf} were taken before this change, with each module's
original middleware. \emph{Residual risk.} The defect was found while composing; it was not introduced by
composition alone, since the services' own middleware already differ. The start-up check confirms that a middleware
is named, not that it is mounted (F07), and matching behaviour between the editions does not by itself show that
either isolates tenants correctly; the cases below test that directly.

\textbf{Cross-tenant isolation and write authorisation (E3).} Because a matching response is not proof that tenant
isolation or authorisation worked, eleven targeted cases were run on each edition after the workload, each from a
fresh deployment (Table~\ref{tab:xt}). A second tenant was added, copied from the first with its bank parameters,
and one internal account was cloned into it; two further users were added to the first tenant, one whose role holds
only the customer-management permissions and one whose role holds every transaction permission except
\code{AUTHORIZE\_TXN}. All eleven cases met their expectation in both editions, with the same status codes and the
same persisted legs. The sweep under the second tenant's token (X-R3) checks one marker, the first tenant's account
numbers, in every response body; it does not prove the absence of every other kind of cross-tenant data. In X-W1 the
refusal comes from the limit check, because under the second tenant the first tenant's user and its limits are not
visible; account isolation on the write path is shown by X-W2.

\begin{table}[t]
\centering\scriptsize
\caption{Cross-tenant and write-authorisation cases; identical outcome in both editions, each from a fresh deployment.}
\label{tab:xt}
\setlength{\tabcolsep}{3pt}
\begin{tabularx}{\columnwidth}{@{}lL>{\raggedright\arraybackslash}p{3.2cm}@{}}
\toprule
Case & What is attempted & Observed (both editions)\\
\midrule
X-R1 & tenant A lists internal accounts & A's 6, none of B's\\
X-R2 & tenant B lists internal accounts & B's 1, none of A's\\
X-R3 & tenant-B token on all 231 parameterless GETs & 114 answer 200; no A account number in any body\\
X-W1 & tenant-B token posts between two of A's accounts & 400 (limit check); nothing persisted\\
X-W2 & tenant A posts to B's account & 400 account not found; nothing persisted\\
A-W1 & post without a token & 401; nothing persisted\\
A-W2 & post, role without transaction permissions & 403; nothing persisted\\
A-W3 & save, same role & 403; nothing persisted\\
A-W4 & authorise a saved transaction, same role & 403; still Entered\\
A-W5 & authorise, every transaction permission except \code{AUTHORIZE\_TXN} & 403; still Entered\\
A-W6 & control: the checker authorises the same transaction & 200; posted\\
\bottomrule
\end{tabularx}
\end{table}

\textbf{Answer to RQ1.} The composed edition has route parity on all 844 operations (a statement about methods and paths, not about request
and response contracts), and within the tested coverage
(47 workload cases, 231 read paths under five principals, and the eleven cross-tenant and write-authorisation cases)
it is behaviourally equivalent to the nine-service edition. The equivalence holds for these inputs and under the
normalisation rules of Section~\ref{sec:design}, not beyond them. It is not claimed for paths this study did not
exercise: the end-of-day run, bulk posting, standing instructions, the channels module past authorisation, writes
under other principals beyond Table~\ref{tab:xt}, parameterised reads, and queue-driven processing.

% =====================================================================================================
\section{RQ2: Detection of Injected Composition Faults}
\label{sec:rq2}

\subsection{The edition's check suite}
\label{sec:checks}

\begin{figure}[t]
\centering\begin{tikzpicture}[node distance=2mm]
\node[fb=7.2cm, cN] (s1) {\textbf{Before the build:} verify-layout and doctor: sibling checkouts, branches, SDK, feed token};
\node[fb=7.2cm, cN, below=1.6mm of s1] (s2) {\textbf{Restore:} check\_pins: all 195 shared-package pins agree (check-latest-versions reports, never edits)};
\draw[ar] (s1) -- (s2);
\node[fb=7.2cm, cB, below=1.6mm of s2] (s3) {\textbf{Build:} the version-conflict warning (MSB3277) is an error in this repository};
\draw[ar] (s2) -- (s3);
\node[fb=7.2cm, cB, below=1.6mm of s3] (s4) {\textbf{Publish and images:} duplicate output paths surface only here};
\draw[ar] (s3) -- (s4);
\node[fb=7.2cm, cG, below=1.6mm of s4] (s5) {\textbf{Start-up:} fail fast on missing configuration; module-registry, route and scoped-filter assertions};
\draw[ar] (s4) -- (s5);
\node[fb=7.2cm, cG, below=1.6mm of s5] (s6) {\textbf{Route table:} route-parity and diff\_routes: composed OpenAPI document against each standalone document};
\draw[ar] (s5) -- (s6);
\node[fb=7.2cm, cG, below=1.6mm of s6] (s7) {\textbf{Deployed edition:} smoke-deployed and compare-deployed: dependencies, timing, and the standalone control};
\draw[ar] (s6) -- (s7);
\node[fb=7.2cm, cR, below=1.6mm of s7] (s8) {\textbf{Around changes:} check\_corrections before and after every upstream merge; check\_migration\_drift before migrating an existing database};
\draw[ar] (s7) -- (s8);
\end{tikzpicture}
\caption{Where each check sits between a clean checkout and a running host.}
\label{fig:checks}
\end{figure}
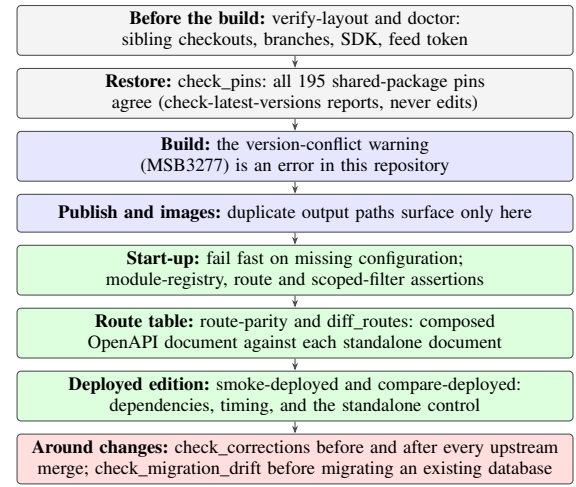

Each check of the edition exists because of a failure that did not announce itself, and the checks sit at different
points between a clean checkout and a running host (Fig.~\ref{fig:checks}, Table~\ref{tab:scripts}). The scripts share
a design: they are read-only, so running one never changes the system under test; they report every problem and not
only the first; and they exit non-zero on failure, so each can be a release gate.

\begin{table}[t]
\centering\footnotesize
\caption{The edition's check scripts and the failure each guards against; the ledger validator is this study's.}
\label{tab:scripts}
\begin{tabularx}{\columnwidth}{@{}lL@{}}
\toprule
Script & Guards against\\
\midrule
verify-layout, doctor & a missing sibling checkout, a wrong branch, a missing SDK or feed token, each of which otherwise looks like an unrelated build error\\
check\_pins & one stale shared-package pin, a run-time type-load failure and not a build error\\
check-latest-versions & silent version drift; it reports and never edits, because a bump is a decision\\
check\_corrections & a fix silently removed by an upstream merge (18 entries)\\
check\_migration\_drift & a migration that stops part-way because an index name has drifted\\
route-parity, diff\_routes & a missing or extra operation, compared by method and path\\
compare-deployed & failures that occur only in the edition, judged against the standalone service\\
smoke-deployed & a deployment that starts but cannot reach its dependencies, or answers slowly\\
ledger check (this study) & an unbalanced posting, a balance that does not move by its amount, a broken chain, a duplicate, a reversal that does not net to zero\\
\bottomrule
\end{tabularx}
\end{table}

\textbf{Layout and prerequisites.} Because the edition builds against seven sibling checkouts, a first build can fail
for several unrelated reasons, and the SDK's message rarely names the real one: a missing sibling looks like a bad
project reference and a missing feed credential like an authorisation failure on an unrelated package. The
environment check reports each problem with its remedy, and the layout check reports the branch and commit each
repository was built against; a wrong branch is a failure, because it would build cleanly into an image without the
fixes. A second script asks the package registry for the latest stable release of every pin and reports, but never
edits, because raising a package above the version a module was compiled against is the specific risk of this design.

\begin{figure}[t]
\centering\begin{tikzpicture}[node distance=2mm]
\node[fb=7.0cm, cG] (s1) {1. Run the check before the merge: all corrections present, the baseline is green};
\node[fb=7.0cm, cN, below=2.2mm of s1] (s2) {2. Merge from upstream: version control merges non-conflicting changes by itself};
\draw[ar] (s1) -- (s2);
\node[fb=7.0cm, cR, below=2.2mm of s2] (s3) {3. A hunk that removes a correction merges with no conflict: nothing fails; the next sign is somewhere unrelated, or none};
\draw[ar] (s2) -- (s3);
\node[fb=7.0cm, cB, below=2.2mm of s3] (s4) {4. Run the check after the merge: \texttt{GONE <label>} for each missing entry; exit non-zero};
\draw[ar] (s3) -- (s4);
\node[fb=7.4cm, cY, below=3mm of s4] {Each entry: a literal string that must appear (or, with a \texttt{!} prefix, must not appear) in a named file of a named repository. Add one whenever a service change is relied on.};

\end{tikzpicture}
\caption{The correction check around an upstream merge: a fix can disappear without a conflict.}
\label{fig:corrections}
\end{figure}
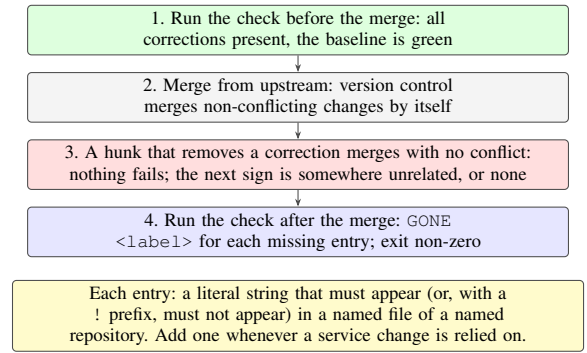

\textbf{Corrections.} The edition composes seven repositories that keep moving, and several carry small changes it
needs. Version control merges non-conflicting changes automatically, so an upstream merge can remove one without a
conflict (Fig.~\ref{fig:corrections}). The correction check holds 18 literal strings that must appear, or with a
prefix must not appear, in named files, in six groups: the tenant-aware business date in the shared library (2), the
registration extensions that make composition possible (8), the model-assembly rename (1), fixes found by running the
hosts (4), fixes found by testing a deployed edition (2), and a host-owned concern that must stay out of a module
extension (1). One deployed-edition fix shows why layered configuration needs a control: a path that lacked its
leading slash was joined to a base address with no separator, so an endpoint answered 500 in the composed API and 200
standalone. As Section~\ref{sec:rq2} reports, the script as committed works only on Windows; a verified one-line patch fixes it.

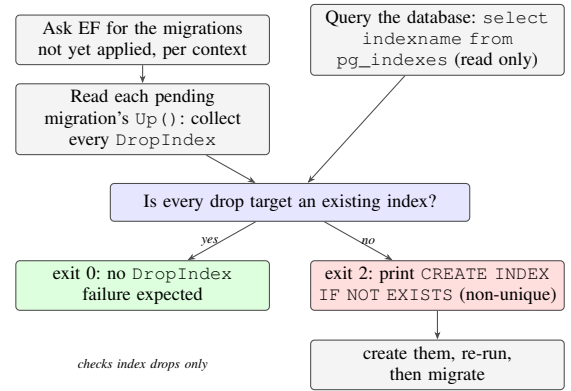
\begin{figure}[t]
\centering\begin{tikzpicture}[node distance=2mm]

\node[fb=3.2cm, cN] (ef) at (-1.95,0) {Ask EF for the migrations not yet applied, per context};
\node[fb=3.2cm, cN] (q) at (1.95,0) {Query the database: \texttt{select indexname from pg\_indexes} (read only)};
\node[fb=3.2cm, cN] (up) at (-1.95,-1.05) {Read each pending migration's \texttt{Up()}: collect every \texttt{DropIndex}};
\draw[ar] (ef) -- (up);
\node[fb=4.6cm, cB] (d) at (0,-2.15) {Is every drop target an existing index?};
\draw[ar] (up) -- (d); \draw[ar] (q) -- (d);
\node[fb=3.2cm, cG] (y) at (-1.95,-3.25) {exit 0: no \texttt{DropIndex} failure expected};
\node[fb=3.2cm, cR] (n) at (1.95,-3.25) {exit 2: print \texttt{CREATE INDEX IF NOT EXISTS} (non-unique)};
\draw[ar] (d) -- node[lab, left] {yes} (y); \draw[ar] (d) -- node[lab, right] {no} (n);
\node[fb=3.2cm, cN] (fix) at (1.95,-4.3) {create them, re-run, then migrate};
\draw[ar] (n) -- (fix);
\node[lab] at (-1.95,-4.3) {checks index drops only};

\end{tikzpicture}
\caption{The migration-drift pre-flight: read-only, and it names the indexes to create first.}
\label{fig:drift}
\end{figure}

\textbf{Migration drift.} The drift script (Fig.~\ref{fig:drift}) reads the migration source and runs one query for
the index names, never writing. It exits 0 when every drop target exists, 2 when some are missing and 1 on an error,
and it prints statements that create the missing indexes first, non-unique, because uniqueness is irrelevant to a drop
and a unique index could fail on data that already has duplicates. By default it examines only migrations not yet
applied, and it checks nothing other than index drops.

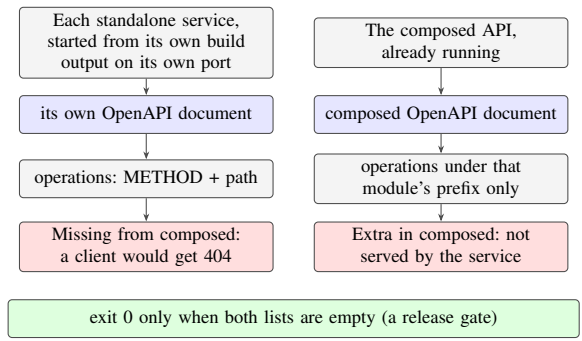
\begin{figure}[t]
\centering\begin{tikzpicture}[node distance=2mm]

\foreach \c/\x/\a/\b/\cc/\d in {L/-1.95/{Each standalone service, started from its own build output on its own port}/{its own OpenAPI document}/{operations: METHOD + path}/{Missing from composed: a client would get 404},
                                R/1.95/{The composed API, already running}/{composed OpenAPI document}/{operations under that module's prefix only}/{Extra in composed: not served by the service}} {
  \node[fb=3.2cm, cN] (\c1) at (\x,0) {\a};
  \node[fb=3.2cm, cB] (\c2) at (\x,-0.95) {\b};
  \node[fb=3.2cm, cN] (\c3) at (\x,-1.8) {\cc};
  \node[fb=3.2cm, cR] (\c4) at (\x,-2.7) {\d};
  \draw[ar] (\c1) -- (\c2); \draw[ar] (\c2) -- (\c3); \draw[ar] (\c3) -- (\c4);
}
\node[fb=7.4cm, cG] at (0,-3.65) {exit 0 only when both lists are empty (a release gate)};

\end{tikzpicture}
\caption{The route-parity check: operations of each standalone service against the composed document.}
\label{fig:parity}
\end{figure}

\textbf{Route parity.} The parity script (Fig.~\ref{fig:parity}) starts each service from its own build output,
fetches the OpenAPI document the service generates for itself, and diffs it against the composed document under that
module's prefix. Comparing generated documents rather than route attributes means that a convention, or an
application part that pulls in an extra controller, is caught and not reasoned about (it caught F09). It compares
operations rather than paths, because a path that serves GET on one side and GET and POST on the other is a
difference a client would meet. Two operational details were recorded in it: a prefix passed with a leading slash was
rewritten by a Windows shell into a file path, which made every match fail, and a trailing carriage return made the
request tool reject every address, which read as the whole API being down.

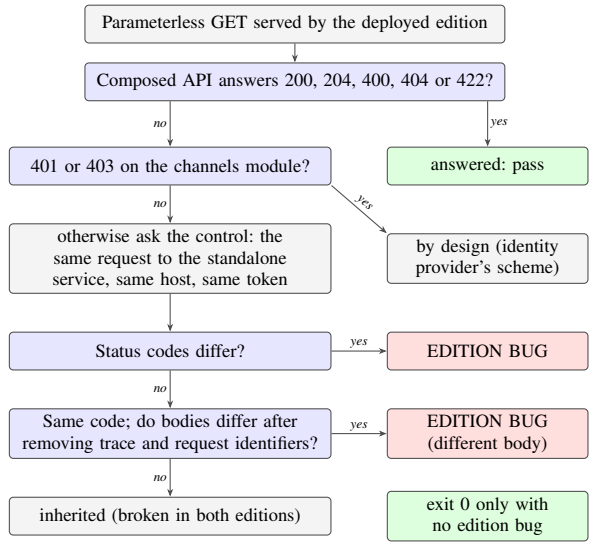
\begin{figure}[t]
\centering\begin{tikzpicture}[node distance=2mm]

\node[fb=5.4cm, cN] (g) at (0,0) {Parameterless GET served by the deployed edition};
\node[fb=5.4cm, cB, below=2.5mm of g] (a) {Composed API answers 200, 204, 400, 404 or 422?};
\draw[ar] (g) -- (a);
\node[fb=4.1cm, cB, anchor=north] (ch) at ($(a.south)+(-1.65,-6mm)$) {401 or 403 on the channels module?};
\path (2.55,0) coordinate (rc);
\node[fb=2.5cm, cG] (ok) at (rc |- ch) {answered: pass};
\draw[ar] (a.south -| ok.north) -- node[lab, right] {yes} (ok.north);
\draw[ar] (a.south -| ch) -- node[lab, left] {no} (ch);
\node[fb=4.1cm, cN, below=5mm of ch] (ctl) {otherwise ask the control: the same request to the standalone service, same host, same token};
\node[fb=2.5cm, cN] (des) at (rc |- ctl) {by design (identity provider's scheme)};
\draw[ar] (ch.south east) -- node[lab, above, sloped] {yes} (des.north west);
\draw[ar] (ch) -- node[lab, left] {no} (ctl);
\node[fb=4.1cm, cB, below=5mm of ctl] (st) {Status codes differ?};
\node[fb=2.5cm, cR] (b1) at (rc |- st) {EDITION BUG};
\draw[ar] (ctl) -- (st); \draw[ar] (st) -- node[lab, above] {yes} (b1);
\node[fb=4.1cm, cB, below=5mm of st] (bd) {Same code; do bodies differ after removing trace and request identifiers?};
\node[fb=2.5cm, cR] (b2) at (rc |- bd) {EDITION BUG (different body)};
\draw[ar] (st) -- node[lab, left] {no} (bd); \draw[ar] (bd) -- node[lab, above] {yes} (b2);
\node[fb=4.1cm, cN, below=5mm of bd] (inh) {inherited (broken in both editions)};
\node[fb=2.5cm, cG] (ex) at (rc |- inh) {exit 0 only with no edition bug};
\draw[ar] (bd) -- node[lab, left] {no} (inh);

\end{tikzpicture}
\caption{The deployed comparison: each failing endpoint is judged against the standalone service.}
\label{fig:compare}
\end{figure}

\textbf{The deployed comparison.} This script (Fig.~\ref{fig:compare}) probes every parameterless GET that a deployed
edition serves and, for each that does not succeed, asks whether the standalone service does better, on the same
host with the same token. A 500 in the composed API means little on its own, because some endpoints are broken in both
editions; the same 500 where the standalone service answers 200 is an edition bug. An earlier sweep had probed 25
endpoints per module and reported no composition-specific error, yet one module serves 89 parameterless GETs and five
that failed only in the edition were outside the sample, so the script has no cap. The comparison covers bodies as well as status codes. An endpoint that answered 500 in both
editions had been called inherited, until the bodies showed that the composed host returned an empty response where
the standalone service returned a structured error, because the module's exception middleware was not mounted; the host
now mounts each module's exception middleware. In this study all 231 parameterless GETs were compared across the two
editions, with no cap (88 in the accounts module): none failed only in the composed edition, and every endpoint that
returned an error in both returned the same body, so neither the five failures the sample had missed nor the empty
error response is still present. The read probes of this study (Section~\ref{sec:design}) extend the same idea to five
principals.

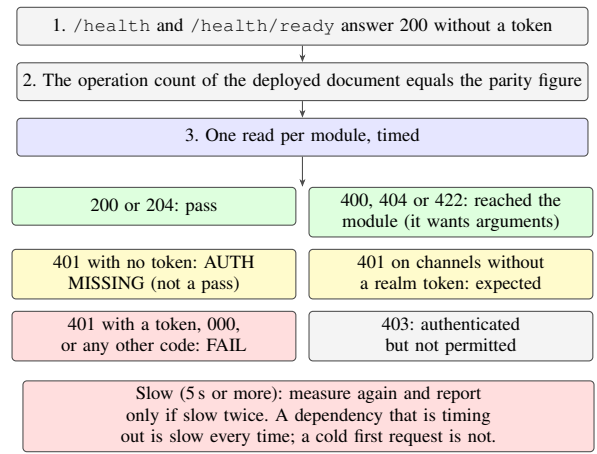
\begin{figure}[t]
\centering\begin{tikzpicture}[node distance=2mm]
\node[fb=7.4cm, cN] (s1) {1. \texttt{/health} and \texttt{/health/ready} answer 200 without a token};
\node[fb=7.4cm, cN, below=2.2mm of s1] (s2) {2. The operation count of the deployed document equals the parity figure};
\draw[ar] (s1) -- (s2);
\node[fb=7.4cm, cB, below=2.2mm of s2] (s3) {3. One read per module, timed};
\draw[ar] (s2) -- (s3);
\node[fb=3.6cm, cG, anchor=north east] (r1) at ($(s3.south)+(-0.08,-4mm)$) {200 or 204: pass};
\node[fb=3.6cm, cG, anchor=north west] (r2) at ($(s3.south)+(0.08,-4mm)$) {400, 404 or 422: reached the module (it wants arguments)};
\draw[ar] (s3.south) -- ++(0,-4mm);
\node[fb=3.6cm, cY, anchor=north east] (r3) at ($(r1.south east |- r2.south)+(0,-1.5mm)$) {401 with no token: AUTH MISSING (not a pass)};
\node[fb=3.6cm, cY, anchor=north west] (r4) at ($(r2.south west)+(0,-1.5mm)$) {401 on channels without a realm token: expected};
\node[fb=3.6cm, cR, anchor=north east] (r5) at ($(r3.south east |- r4.south)+(0,-1.5mm)$) {401 with a token, 000, or any other code: FAIL};
\node[fb=3.6cm, cN, anchor=north west] (r6) at ($(r4.south west)+(0,-1.5mm)$) {403: authenticated but not permitted};
\node[fb=7.4cm, cR, anchor=north] at ($(r6.south west)+(0,-2.5mm)$) {Slow (5\,s or more): measure again and report only if slow twice. A dependency that is timing out is slow every time; a cold first request is not.};

\end{tikzpicture}
\caption{The deployed smoke test: how each response is classified.}
\label{fig:smoke}
\end{figure}

\textbf{The deployed smoke test.} Route parity proves that the route table is the same and an image test that a
container starts, but neither proves that a deployed task reaches its database, resolves its secrets and returns data
for every module. The smoke test (Fig.~\ref{fig:smoke}) checks that liveness and readiness answer without a token, that
the operation count of the deployed document equals the parity figure, and that one read per module succeeds. It
reports a missing token as authorisation missing, not as a pass, and it measures time, because a dependency that
cannot be reached usually waits rather than fails: an unreachable audit store once made every request that writes an
audit record take about 30\,s and still answer 200. A slow first result is measured again, because a first request on a
cold task pays for compilation, model building and connection warm-up (N3).

\begin{figure}[t]
\centering\begin{tikzpicture}[node distance=2mm]

\node[fb=3.4cm, cN] (i1) at (-1.95,0) {Ledger legs (PostgreSQL)};
\node[fb=3.4cm, cN] (i2) at (1.95,0) {Account balances (PostgreSQL)};
\node[fb=7.4cm, cB, below=4mm of {$(i1)!0.5!(i2)$}] (s1) {1. Per transaction: debits = credits; legs numbered 1..n};
\node[fb=7.4cm, cB, below=1.6mm of s1] (s2) {2. Per leg: the balance moves by exactly its amount};
\draw[ar] (s1) -- (s2);
\node[fb=7.4cm, cB, below=1.6mm of s2] (s3) {3. Per account: legs chain by balance from zero (each leg starts where another ended)};
\draw[ar] (s2) -- (s3);
\node[fb=7.4cm, cB, below=1.6mm of s3] (s4) {4. Per account: the clear balance equals the end of the chain};
\draw[ar] (s3) -- (s4);
\node[fb=7.4cm, cB, below=1.6mm of s4] (s5) {5. Reversals: the original and its reversal net to zero per account};
\draw[ar] (s4) -- (s5);
\node[fb=7.4cm, cB, below=1.6mm of s5] (s6) {6. Duplicate candidates: transactions with identical legs (account, side, amount, narration)};
\draw[ar] (s5) -- (s6);
\draw[ar] (i1) -- (s1.north -| i1); \draw[ar] (i2) -- (s1.north -| i2);
\node[fb=2.3cm, cG, below=2.5mm of s6.south west, anchor=north west] {no finding};
\node[fb=2.3cm, cY, below=2.5mm of s6.south] {duplicate candidates listed};
\node[fb=2.3cm, cR, below=2.5mm of s6.south east, anchor=north east] {finding: exit 1};

\end{tikzpicture}
\caption{The ledger check: groups of checks over the ledger legs, with three outcomes.}
\label{fig:ledger}
\end{figure}
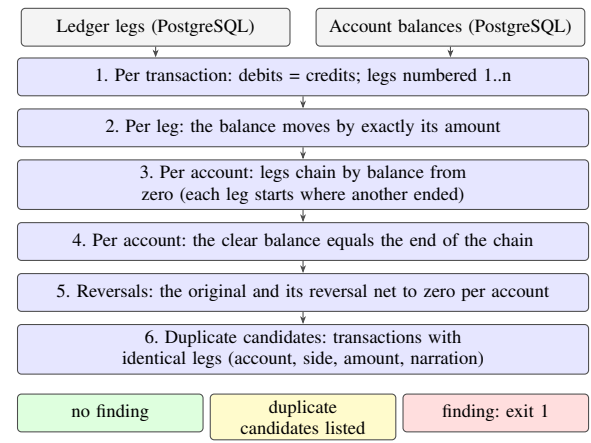

\textbf{The ledger check.} None of the edition's scripts reads a ledger. The ledger check of this study
(Fig.~\ref{fig:ledger}) applies the double-entry rules to the persisted legs with exact decimals, reports every finding,
and names accounts instead of printing their numbers. It was first tested on six small synthetic ledgers, one per outcome (good,
unbalanced, half-posted, duplicated, correct reversal, wrong-amount reversal), each giving the expected exit code, and
in this study it checked every ledger that the workload runs and the crash tests produced.

\textbf{The images (E6).} For the earlier report, all three images were built and run against PostgreSQL, MongoDB,
Redis and a queue endpoint: inside the API image all ten module settings files were loaded, the actions were served
with no duplicate routes, both scoped filters stayed inside their modules, and tenant-scoped data was returned for
accounts, customers and transactions; the gateway container routed into the API container. The image history showed
no credential in any layer and no package-source file. Images were 723\,MB (API), 693\,MB (consumer) and 586\,MB
(background), starting in 1 to 3\,s in the container. An ordinary build cannot validate the images: a
duplicate-output-path error covering 31 paths went undetected through eight phases of work and surfaced the first time
anything was published, because a build filters such collisions and only a publish computes the files to ship.

\subsection{Fault injection}

\textbf{Fault set.} The author designed 16 faults before running them, to cover the failure classes the earlier
report names as the reason for each check (dependency pins, merges that drop a correction, module ordering,
duplicate routes, scoped filters, current-user middleware, dropped controllers, consumer queues, missing settings),
plus classes it does not name: configuration that is present but wrong (F15, F16), and a change of service business
logic (F11). Each was injected alone into copies of the repositories or the hosts'
configuration. The edition's detectors then ran in release order: the pin check, the correction
check, the build, start-up of the affected host, and route parity. A fault that passed all of them was run through
the behavioural harness of Section~\ref{sec:design} on the full deployment and compared with a clean baseline case
by case and endpoint by endpoint. Each change was reverted and the source tree verified against a hash.

\begin{table*}[t]
\centering\footnotesize
\caption{Fault injection. Detectors: P pins, C corrections, B build, S start-up assertion, R route parity (the
edition's suite); W workload and probes (this study). \yes{} detected, -- not detected, blank: not reached.}
\label{tab:faults}
\begin{tabularx}{\textwidth}{@{}clLcccccc L@{}}
\toprule
& Class & Injected fault & P & C & B & S & R & W & How it showed\\
\midrule
F01 & dependency & one repository pins a shared package at 1.0.9 & \yes & -- & & & & & pin disagreement\\
F02 & merge & upstream merge removes the Customers caching registration & -- & \yes & -- & -- & -- & & correction gone\\
F03 & composition & module list reordered, configuration order not & -- & -- & -- & \yes & & & host refuses to start\\
F04 & composition & host controller collides with a module route & -- & -- & -- & \yes & & & duplicate route at start-up\\
F05 & composition & permission filter leaks to every module & -- & -- & -- & \yes & & & filter on 773 foreign actions\\
F06 & composition & transactions module's scoped filter list emptied & -- & -- & -- & -- & -- & \yes & one probe: 403 $\rightarrow$ 200 for a user with no role\\
F07 & composition & current-user middleware not mounted for one module & -- & -- & -- & -- & -- & \yes & every posting 400\\
F08 & composition & exception and current-user middleware swapped & -- & -- & -- & -- & -- & \yes & tenantless requests 422 $\rightarrow$ 500\\
F09 & composition & one module controller dropped from the host & -- & -- & -- & -- & \yes & & operations missing\\
F10 & merge & audit sink re-registered inside a module extension & -- & \yes & -- & -- & -- & & forbidden string present\\
F11 & service logic & authorise no longer requires status ``Entered'' & -- & -- & -- & -- & -- & -- & not detected\\
F12 & configuration & two consumer workers on one queue & -- & -- & & \yes & & & host refuses to start\\
F13 & configuration & a consumer worker's queue name missing & -- & -- & & \yes & & & host refuses to start\\
F14 & configuration & cache endpoint missing & -- & -- & & \yes & & & host refuses to start\\
F15 & configuration & token audience differs from the issuer's & -- & -- & & -- & -- & \yes & every request 401\\
F16 & configuration & database connection points at another database & -- & -- & & -- & -- & \yes & authorisation 403, approvals fail\\
\midrule
& & \textbf{Detected (of the faults reaching the detector)} & 1 & 2 & 0 & 6 & 1 & 5 of 6 & suite 10 of 16; with W 15 of 16\\
\bottomrule
\end{tabularx}
\end{table*}

\begin{table}[t]
\centering\footnotesize
\caption{Detection of the 16 injected faults by class.}
\label{tab:faultclass}
\begin{tabular}{@{}lccc@{}}
\toprule
Class & Faults & Suite & Suite + W\\
\midrule
Composition (F03--F09) & 7 & 4 & 7\\
Dependency and merge (F01, F02, F10) & 3 & 3 & 3\\
Configuration (F12--F16) & 5 & 3 & 5\\
Service logic (F11) & 1 & 0 & 0\\
\midrule
All & 16 & 10 & 15\\
\bottomrule
\end{tabular}
\end{table}

\textbf{Results.} Table~\ref{tab:faults} gives each fault and Table~\ref{tab:faultclass} the counts by class. The
suite detects 10 of these 16 faults, each at the stage it was designed for: start-up assertions stop six with a
message that names the cause, the pin and correction checks catch dependency and merge faults before a build, and
route parity catches the dropped controller. Of the six it misses, three are composition faults that leave the
route table and the registry's invariants intact but change what runs per request (F06--F08); F06 is subtle,
because the scoped-filter assertion checks only the filters a module declares, so a module that stops declaring its
filter passes. Two are configuration values that are present but wrong (F15, F16): the host checks that the token
settings and connection string exist, not that they work, and its readiness probe checks only that PostgreSQL
answers. One is a change of business logic in a service (F11).
The behavioural harness detects five of the six, so of the 16 faults the suite catches 10, the behavioural harness
five more, and one (F11) is caught by neither. F06 is the narrowest catch, on one of 231 probed paths. F11 is
detected by nothing: the workload's re-authorise case was still rejected with the same status, so the weakened rule
never changed an observed outcome.

\textbf{Interpretation.} These counts describe this fault set, not the probability of detecting an arbitrary future
defect. The faults were chosen by the author, who also wrote the behavioural harness; the classes the edition's
checks were written for are the ones they catch, and the classes added for this study are the ones they miss, so
the set may favour both the suite on its own classes and the harness on the new ones. One instrument problem also
matters: the correction check names its files with Windows path separators, so on Linux or macOS it reports all 18
corrections missing on a clean tree, and a Linux release gate would fail on every run. A one-line fix (splitting each
entry on the Windows separator before joining it) makes it portable. With the fix, a clean
tree reports only the two entries that live in the shared library's repository, which this study did not check out,
and the injected faults F02 and F10 are reported as \code{GONE}; the results in Table~\ref{tab:faults} were obtained
with the same fix.

\textbf{Answer to RQ2.} On this fault set the suite catches assembly faults that its checks were written for (10 of
16) and misses faults that change what a correctly registered host does per request, configuration that is present
but wrong, and a change of service business logic. Behavioural tests on the running system with several
principals caught five of those six; the sixth, a change of service business logic (F11), was caught by neither.

% =====================================================================================================
\section{RQ3: Failure Behaviour (Emulated)}
\label{sec:rq3fail}

\FailSection

% =====================================================================================================
\section{RQ3: Performance and Cost (Emulated)}
\label{sec:rq3perf}

\PerfSection

% =====================================================================================================
\section{Findings}
\label{sec:findings}

Table~\ref{tab:findings} collects what the experiments exposed. The first group concerns the edition, the second
the platform that both editions share, and the third the nine-service edition alone. The shared defects are part of
the behaviour the edition reproduces faithfully; a differential evaluation finds them because the expected outcomes
of the workload and the five principals of the probes were judged against intended behaviour as well as against the
other edition.

\begin{table*}[t]
\centering\footnotesize
\caption{Findings. Severity is the author's judgement; none of these defects is claimed to be fixed.}
\label{tab:findings}
\begin{tabularx}{\textwidth}{@{}clLlL@{}}
\toprule
& Severity & Finding & Scope & Evidence and suggested action\\
\midrule
\multicolumn{5}{@{}l}{\emph{The edition and its checks}}\\
X1 & medium & One crash takes all five modules down & edition & Table~\ref{tab:kill}; run two API tasks, or keep the posting module separate\\
X2 & medium & Six of 16 injected faults pass the check suite & edition & Table~\ref{tab:faults}; add multi-principal behavioural probes as a release gate\\
X3 & medium & Readiness checks PostgreSQL only; the earlier report said PostgreSQL, MongoDB and Redis & edition & host source; F16 passed readiness; see Section~\ref{sec:discussion}\\
X4 & low & Correction check works only with Windows paths & edition & one-line patch, verified\\
X5 & low & Users-module PDF generation is disabled in the consumer host & edition & the standalone consumer instead downloads a browser at start-up, which needs internet egress on every cold start\\
X6 & medium & Consumer host uses \ConsIdleNew\% of a core while idle & edition & Table~\ref{tab:resources}; poll with a wait time\\
\midrule
\multicolumn{5}{@{}l}{\emph{The platform (both editions)}}\\
P1 & note & Segregation of duties is enforced through roles: authorising needs \code{AUTHORIZE\_TXN}, and a user who holds both the save and the authorise permission may authorise their own transaction by design & both & workload, source; banks wanting four-eyes approval for all users assign the two permissions to different roles\\
P2 & medium & SNS publish inside the account lock: an AWS stall blocks all postings to the accounts involved & both & outage test; publish after commit through an outbox\\
P3 & medium & Health endpoints stay 200 when AWS is unreachable & both & outage test; see Section~\ref{sec:discussion}\\
P4 & low & Gateway circuit breaker effectively disabled (99,999 exceptions before breaking) & both & gateway configuration\\
P5 & medium & Users-module queue client registered as scoped but resolved from the root provider: the poller dies when scope validation is on, and works by accident when it is off & both & source, development run; register the client as a singleton or resolve it inside a scope; not fixed\\
\midrule
\multicolumn{5}{@{}l}{\emph{The nine-service edition}}\\
N1 & medium & Accounts service exceeds its 512\,MiB task under load (up to 524\,MiB, 6 of 10 rounds) & nine-service & Table~\ref{tab:resources}; raise to 1\,GiB\\
N2 & medium & Transaction consumer uses its full 0.25 vCPU while idle & nine-service & Table~\ref{tab:resources}; poll with a wait time\\
N3 & low & At 0.25 vCPU a service's first request after idle takes about 13\,s & nine-service & Table~\ref{tab:kill}; warm on start or size up\\
\bottomrule
\end{tabularx}
\end{table*}

None of P1--P5 is caused by composition, and none is fixed by it. Each was observed on the emulated deployment
built from the same commits as the deployed platform. Reporting them here does not replace disclosure to the platform's owner, which has preceded publication.

\textbf{Defects found while composing (E6).} Before this study, running the composed host exposed four defects that
no build could fail. The users module had no identity framework, because the users service's database registration
bundles a host-owned part with a module-owned part, and leaving that method to the host dropped the identity
registration, so six service descriptors could not resolve; this was visible only because the development
environment validates the service provider at build. No current-user middleware was registered at all, and the
consequence was not an error: with no tenant set, the shared global tenant filter does not apply, so requests ran
without tenant scoping and still answered 200; it is now mounted for every module, one implementation shared by the
four token-based modules, and its declaration is checked at start-up. The OpenAPI
document failed to generate, because one type name exists in two namespaces and the default schema identifier is the
bare type name; one collision means no document, not a partial one, so identifiers are now namespace-qualified. And
the queue library's message pump resolves the queue client from the root provider while the users module registers
it as scoped, so in the development environment the poller dies; the standalone users API does the same, in
production scope validation is off and it works by accident, and the host survives it because it ignores background
failures.

% =====================================================================================================
\section{Architectural Decisions, Trade-offs and Risks}
\label{sec:decisions}

Table~\ref{tab:adr} records the main decisions in the form of an architecture decision record \cite{nygard2011adr}:
what was chosen, over what, and what follows, now with the evidence from this study.

\begin{table*}[t]
\centering\footnotesize
\caption{Architectural decisions, their consequences, and what this study measured about them.}
\label{tab:adr}
\begin{tabularx}{\textwidth}{@{}>{\raggedright\arraybackslash}p{3.3cm}>{\raggedright\arraybackslash}p{2.6cm}>{\raggedright\arraybackslash}p{4.3cm}L@{}}
\toprule
Decision & Instead of & Consequence & Evidence from this study\\
\midrule
Compose by reference & copying service code & no fork; needs sibling checkouts on expected branches & route parity 844/844; new accounts operations reached both editions unedited\\
Explicit module manifest & scanning for modules & order visible; duplicates asserted & F03, F04 stopped at start-up\\
Host owns logging, health, auth, contexts, audit, cache & each module its own & one of each; modules must not register them & F14 stopped at start-up; F15, F16 (present but wrong) did not\\
Separate API, consumer, background hosts & one process & the consumer needs no load balancer; EOD isolated from API restarts & F12, F13 stopped the consumer host\\
Declared configuration precedence & undefined layering & later module wins; conflicts logged & 12 staging and 10 production keys disagree between modules\\
Version ceiling at what binaries bind & latest of everything & no type-load failures; advisories remain & F01 caught by the pin check\\
No schema change & edition-owned migrations & existing databases usable & both editions on one schema, identical ledgers\\
One current-user middleware for token-based modules & each module its own & one strict tenant check; channels keeps its own scheme & repeated workload identical; no-tenant requests refused on all 231 paths\\
Fail fast at start-up & tolerating bad composition & a bad composition cannot boot & 6 of 16 faults stopped with a message naming the cause\\
Ignore background failures & crash the host & a partial failure replaces a total one & health stays 200 while AWS is unreachable (P3)\\
One API task & several API tasks & one crash stops all modules & every module down 16--22\,s per crash (Table~\ref{tab:kill})\\
\bottomrule
\end{tabularx}
\end{table*}

\textbf{Sensitivity points.} Module order decides which configuration value wins, so reordering the manifest can
change behaviour without a code change. The version of each shared package decides whether a module compiled against
an older release still binds. Any middleware or filter a module registers globally reaches every module in the host.
The CPU of the one API task is shared, which this study found to help: under the real task sizes a busy module borrows
what idle modules leave (Section~\ref{sec:rq3perf}).

\textbf{Trade-off points.} Fewer deployable units improve operability, cost and, at these sizes, performance, but one
process shares one dependency set and one failure domain. Ignoring background-service failures keeps the host
answering, at the price of a failure nobody sees. Raising dependencies reduces exposure to advisories but is limited
by what the pre-compiled binaries can bind against.

\textbf{Risks and their status.} To resolve before production use: the shared defects P2 and P5 of
Table~\ref{tab:findings}; worker-level health, because liveness is dependency-free by design and a process with a
stopped queue worker is not healthy, so each worker needs a readiness or heartbeat signal, an alert on its failure log
and a restart policy; a readiness contract (Section~\ref{sec:discussion}); and the open advisories, whose exposure
should be assessed for the deployment. Acceptable for now, with a check or a log line watching them: module boundaries
kept by convention, and configuration collisions, which the host logs at start-up. This classification is the
author's judgement and has not been reviewed by a security or operations team.

% =====================================================================================================
\section{Status and Earlier Evidence}
\label{sec:status}

The edition was built in eight phases (Table~\ref{tab:phases}); the 205 tests in the six service repositories are the
regression net and keep running there, and the edition does not duplicate them. Two pieces of earlier evidence
complement this study.

\begin{table}[t]
\centering\footnotesize
\caption{Status by phase, as reported for the earlier version (E6).}
\label{tab:phases}
\begin{tabularx}{\columnwidth}{@{}clL@{}}
\toprule
Phase & Work & State\\
\midrule
1 & scaffold, module contract, API host & boots; liveness 200\\
2 & customers module & 33 actions; 40 tests green\\
3 & channels module & 116 actions in total; 26 + 40 tests green\\
4 & transactions module & 203 actions; 25 tests green\\
5 & accounts module & 485 actions; 22 tests green\\
6 & users module & all five composed; 46 tests green\\
7 & consumer host & three workers composed\\
8 & background host, Dockerfiles, pipeline & three hosts and images\\
\bottomrule
\end{tabularx}
\end{table}

\textbf{Write-path evidence from a deployed environment (E6).} After the edition was deployed, the author exported the
ledger rows for three posted transactions and the account rows they touch. The export holds six legs: one funds
transfer received through the channels path and two teller cash deposits, each with a debit leg to an internal ledger
account and a credit leg to a customer account. For every transaction the debits equal the credits, the two legs are
numbered 1 and 2, and each leg's balance moves by exactly its amount; the ledger account debited by the two deposits
carries its balance from one to the next without a gap; and for the two customer accounts in the account export, the
clear balance equals the balance after the transaction and the effective available balance equals the clear balance
less the minimum balance. A separate export held the eight legs of one earlier reversed transaction (four debits and
four credits of 1,010.75 each side), every leg flagged as reversed, but not the legs of the reversal itself. The
export shows that posted writes reach the shared schema as
balanced entries, and the controlled experiments of Sections~\ref{sec:rq1}--\ref{sec:rq3fail} add reversals,
rejections, concurrency, crashes and the comparison between editions that the export could not.

\textbf{Earlier single-machine measurements (E6, superseded).} The earlier report measured the API host on one
Linux machine with no CPU limit, eight clients for 20\,s against empty tables: start to healthy in 1.9--2.1\,s, a
working set of 170\,MB idle and 534\,MB under load, 339 requests a second on five endpoints, and an extra gateway hop
of about 8\%. It noted that 534\,MB is above the 0.5\,GiB of one small task, so the API host could not take the
smallest size. Section~\ref{sec:rq3perf} replaces these figures with measurements of both editions under their
Fargate task sizes; the memory observation stands (peak 530\,MiB median, against a 2\,GiB task).

% =====================================================================================================
\section{Discussion}
\label{sec:discussion}

\textbf{When to choose which edition.} Composition buys fewer deployable units, one target group, one log stream
per host and less allocated compute, and in the emulation better read and posting performance, because idle
modules lend their share of the CPU to busy ones, which nine tasks of a quarter vCPU each cannot do. It costs isolation: one
crash, one memory leak or one bad deployment reaches every module, and one process shares one dependency set. A
second API task would remove most of the blast-radius penalty for about \$36 a month at the prices used here.
The composed edition therefore suits small deployments with modest load, where the modules' release cycles are
compatible and the fixed cost of nine services dominates. It is not automatically preferable where modules must scale
independently, where one module's failure must not reach the others, or where strict module boundaries are a
requirement; those needs favour the nine-service edition or a composition that keeps the posting module apart.

\textbf{What the checks are for.} The edition's checks are good at catching a composition that is assembled wrongly.
Their limit is that a correctly assembled host can still behave wrongly. The behavioural harness that caught five of
the six escapes (the sixth, F11, was caught by nothing) is not specific to this platform: the same writes against both editions with a ledger check, and
the same reads under several principals with normalised bodies.

\textbf{Readiness contract.} The edition's liveness endpoint checks nothing and its readiness endpoint checks only
that PostgreSQL answers, so a host pointed at the wrong database (F16) and a host that cannot reach AWS (P3) both
report ready. Neither edition's documentation states what \emph{ready} should mean. We suggest an explicit contract:
\emph{liveness}, the process can make progress, with no dependency checks so that a slow dependency never restarts a
healthy task; \emph{readiness}, the task can serve its core requests, covering the dependencies without which every
request fails (the configured database and schema, the cache that holds posting locks and permissions, and a valid
token configuration); and \emph{dependency health}, reported separately and alerted on, for dependencies whose loss
should degrade rather than remove the task (the audit store, SNS and SQS). Under such a contract F15 and F16 would
have stopped at readiness, and the AWS outage would have raised an alert while postings were blocked.

\textbf{Decisions revisited.} Ignoring background-service failures keeps the API answering when its queue poller
dies, at the price of a failure nobody sees; the outage test confirms that health stays green while AWS is
unreachable. Fail-fast start-up assertions proved their worth: six injected faults stopped the host with a message
that named the cause.

% =====================================================================================================
\section{Threats to Validity and Limitations}
\label{sec:threats}

\textbf{Construct.} Equivalence is judged on what the workload and probes exercise (Section~\ref{sec:rq1}). Read
probes cover parameterless GETs only, and access-control probes cover reads only. Normalisation masks identifiers,
timestamps, trace identifiers and concurrency stamps; a defect confined to those values would not be seen.

\textbf{Internal.} The emulation is not AWS. CPU quotas are enforced by a user-space controller with CFS semantics
rather than by CFS (validated at 96--97\% of quota under load); networking is loopback, so latencies omit the
network between tasks, the ALB and the data stores; moto models the service APIs, not their latency or throttling;
nginx's failure handling differs from an ALB's. All tasks shared one host with the data stores and the load
generator, so ten rounds on that host limit how far the figures generalise. Performance runs were repeated in ten rounds per edition with fresh deployments, randomised scenario
order and alternating editions, but they remain measurements of the emulation: the comparison between editions is
the result, and the absolute throughput and latency figures are not production AWS measurements. Repeating the
performance runs on AWS at the same task sizes is future work.

\textbf{External.} One platform, one team's conventions, one cloud. The fault set was designed by the author
(Section~\ref{sec:rq2}); faults designed independently, by a second engineer or derived from the platform's own
incident history, would test the detection results more fairly and are future work. Module boundaries are kept by
convention and start-up assertions, not by architecture tests: nothing in the edition checks dependency direction
between modules or forbids one module's code from resolving another's services, and such tests (for example
assembly-dependency rules run in the build) would close a gap that neither the suite nor the behavioural harness
covers. The cost model uses list prices for one region and excludes the gateway, the ALB, data
transfer, NAT, logs, storage and the operational effort that motivates the edition.

\textbf{Reliability.} The study ran on copies of the repositories. The platform's source code is proprietary, and the
study's scripts and raw results are not publicly released at this time (Section~\ref{sec:data}). Account identifiers,
internal host names and credentials found in configuration were excluded from the paper.

% =====================================================================================================
\section{Data Availability}
\label{sec:data}

The study's material is not publicly released at this time. \emph{What can be shared:} the emulation and experiment
scripts (deployment, CPU-quota controller, workload with its 47 cases and expected outcomes, ledger check, read
probes, cross-tenant and write-authorisation probes, and the failure, fault-injection and performance harnesses); the raw per-run results of every experiment
reported here, including all ten performance rounds and five failure repetitions per edition; the exact tool versions;
the statistics script, which regenerates the performance and failure tables from the raw results without the
platform; and the patches described in Sections~\ref{sec:composition} and~\ref{sec:rq2}. \emph{Conditions:} on
request to the author, for verifying or reproducing the results reported here, with the agreement of the platform's
owner, and with account identifiers, internal host names and credentials removed. \emph{What cannot be shared:} the
platform's source code, which is proprietary, its credentials and configuration secrets, and the AWS account
identifier. Reproducing the experiments, as opposed to checking the reported numbers, needs access to the platform's
repositories.

% =====================================================================================================
\section{Conclusion}

The consolidated deployment edition composes five independently developed services into one host by reference, while the
services keep shipping on their own. In an emulation of its AWS deployment it has route parity on all 844 operations,
and within the tested coverage it behaves like the nine services it replaces: the same outcomes on 47 workload
cases, the same responses to 1,155 probes apart from one path the emulation answers differently, the same outcome in eleven cross-tenant and
write-authorisation cases, the same ledger to the cent, and the same consistency through repeated
crashes and an AWS outage (RQ1, RQ3). Its own checks catch the injected faults that change what is assembled (10 of 16), and
miss those that change what a correctly assembled host does; behavioural probes caught five of those six, and one,
a change of service business logic, was caught by neither (RQ2). In the emulation
the composed edition \ConclPerf{}, but it widens the failure domain of a crash to every module. These results do not
establish system-wide equivalence or production performance on AWS: the absolute figures are measurements of the
emulation. Because the evaluation is differential and
judged against intended behaviour as well, it also showed defects that both editions share; those, not
the composition, are the first things to fix.

\bibliographystyle{IEEEtran}
\bibliography{refs}

\appendix
\section{Workload Cases and Environment}

Table~\ref{tab:cases} lists the 47 workload cases with the outcome expected before the runs and the status observed,
which was the same in both editions for every case. Table~\ref{tab:versions} gives the environment, and Tables~\ref{tab:roundsrps} and~\ref{tab:roundsp95} the result of
every performance round.

\begin{table}[h]
\centering\scriptsize
\caption{The 47 workload cases. Observed status identical in both editions.}
\label{tab:cases}
\setlength{\tabcolsep}{3pt}
\begin{tabularx}{\columnwidth}{@{}lLlc@{}}
\toprule
Cases & What each does & Expected & Observed\\
\midrule
A1--A5 & create a transaction limit for one of five types, approve as second user & created, approved & 200\\
A6 & grant the limits to the super-administrator role, approve & granted & 200\\
A7--A8 & grant user-level limits to each of the two users, approve & granted & 200\\
A9--A11 & create three general-ledger accounts (asset, liability, revenue), approve & created, approved & 200\\
A12--A14 & create an internal product per type, approve & created, approved & 200\\
A15--A20 & create six internal accounts (two vaults, two settlement, income, hot), approve & created, approved & 200\\
A21 & list the internal accounts & 6 accounts & 200\\
\midrule
B1--B10 & balanced two-leg posting, amounts 0.01 to 125,000.00 & posted & 200\\
B11 & balanced four-leg posting (one debit; principal, fee and tax credits) & posted & 200\\
B12 & debits 100.00, credits 90.00 & rejected & 422\\
B13 & leg on an unknown account & rejected & 400\\
B14--B15 & zero amount; negative amount & rejected & 400\\
B16 & a single leg & rejected & 400\\
B17 & post new legs under an already posted identifier (a posted transaction can only be reversed) & rejected & refused\\
B18 & send the same two-leg posting twice, without an identifier & two transactions, distinct identifiers & 200 (by design)\\
B19 & reverse B1 & reversed & 200\\
B20 & reverse B1 again & rejected & 422\\
B21 & reverse B11 & reversed & 200\\
B22 & save a transaction for approval & saved pending & 200\\
B23 & authorise it as the second user & posted & 200\\
B24 & authorise it again & rejected & 422\\
B25 & save another and authorise it as its maker (a user holding both permissions) & authorised$^\ddagger$ & 200\\
B26 & 20 concurrent debits of 100.00 from one account & all posted or cleanly rejected & 20$\times$200\\
\bottomrule
\end{tabularx}
\\[2pt]{\scriptsize $^\ddagger$ expectation corrected after the runs: a user holding both permissions may authorise their own transaction (P1).}
\end{table}

\begin{table}[h]
\centering\scriptsize
\caption{Environment of the experiments.}
\label{tab:versions}
\begin{tabularx}{\columnwidth}{@{}lL@{}}
\toprule
Component & Version\\
\midrule
Host & one machine, 10 CPU cores, 16\,GiB of memory\\
.NET & SDK 10.0.301, ASP.NET Core runtime 10.0.9; all 13 hosts target net10.0\\
Data stores & PostgreSQL 16.15, MongoDB 8.0.0, Redis 8.10.2\\
AWS emulation & moto 5.2.4 (server mode, waitress 3.0.2), boto3 1.43.114, Python 3.13.0\\
Load balancer & nginx 1.31.7\\
Load generator & oha 1.16.0\\
Prices & AWS Price List API, eu-west-1, publication of 2026-09-11\\
\bottomrule
\end{tabularx}
\end{table}

\begin{table*}[t]
\centering\scriptsize
\caption{Throughput (requests per second) in every performance round. R8, R32, R64: reads with 8, 32 and 64 clients through the gateway; R32A: 32 clients sent straight to the load balancer; W8, W32: postings with 8 and 32 clients. N: nine-service edition; C: composed edition; First: the edition that ran first in the round.}
\label{tab:roundsrps}
\setlength{\tabcolsep}{3.2pt}
\begin{tabular}{@{}ccrrrrrrrrrrrr@{}}
\toprule
Round & First & \multicolumn{2}{c}{R8} & \multicolumn{2}{c}{R32} & \multicolumn{2}{c}{R64} & \multicolumn{2}{c}{R32A} & \multicolumn{2}{c}{W8} & \multicolumn{2}{c}{W32}\\
 & & N & C & N & C & N & C & N & C & N & C & N & C\\
\midrule
1 & N & 136 & 397 & 166 & 279 & 107 & 351 & 135 & 470 & 11.1 & 16.6 & 11.2 & 20.2\\
2 & C & 160 & 402 & 168 & 382 & 137 & 395 & 203 & 392 & 9.7 & 26.6 & 13.9 & 21.7\\
3 & N & 172 & 471 & 158 & 394 & 132 & 386 & 154 & 410 & 13.9 & 25.1 & 14.7 & 23.4\\
4 & C & 172 & 514 & 151 & 410 & 181 & 401 & 197 & 391 & 13.7 & 25.4 & 15.3 & 24.4\\
5 & N & 147 & 516 & 167 & 434 & 160 & 381 & 197 & 418 & 13.1 & 23.7 & 14.3 & 22.9\\
6 & C & 173 & 370 & 168 & 376 & 160 & 355 & 153 & 472 & 14.0 & 18.0 & 10.0 & 32.4\\
7 & N & 174 & 523 & 152 & 436 & 148 & 419 & 175 & 459 & 15.0 & 18.7 & 11.6 & 27.5\\
8 & C & 167 & 521 & 230 & 336 & 174 & 416 & 190 & 489 & 13.6 & 17.1 & 8.6 & 28.5\\
9 & N & 191 & 360 & 174 & 363 & 176 & 312 & 199 & 364 & 12.6 & 14.0 & 9.4 & 24.7\\
10 & C & 156 & 404 & 133 & 356 & 130 & 289 & 133 & 368 & 11.3 & 20.4 & 13.2 & 24.1\\
\bottomrule
\end{tabular}
\end{table*}
\begin{table*}[t]
\centering\scriptsize
\caption{p95 latency (ms) in every performance round; columns as in Table~\ref{tab:roundsrps}.}
\label{tab:roundsp95}
\setlength{\tabcolsep}{3.2pt}
\begin{tabular}{@{}ccrrrrrrrrrrrr@{}}
\toprule
Round & First & \multicolumn{2}{c}{R8} & \multicolumn{2}{c}{R32} & \multicolumn{2}{c}{R64} & \multicolumn{2}{c}{R32A} & \multicolumn{2}{c}{W8} & \multicolumn{2}{c}{W32}\\
 & & N & C & N & C & N & C & N & C & N & C & N & C\\
\midrule
1 & N & 205 & 65.7 & 745 & 260 & 2,446 & 285 & 987 & 109 & 1,711 & 1,433 & 11,336 & 3,606\\
2 & C & 194 & 61.8 & 781 & 132 & 1,900 & 273 & 649 & 141 & 2,205 & 989 & 4,302 & 3,611\\
3 & N & 189 & 63.4 & 789 & 125 & 2,002 & 279 & 798 & 121 & 1,588 & 1,063 & 4,673 & 3,543\\
4 & C & 192 & 55.1 & 860 & 138 & 1,501 & 278 & 616 & 118 & 1,618 & 1,005 & 4,097 & 3,530\\
5 & N & 197 & 54.5 & 797 & 117 & 1,589 & 287 & 690 & 113 & 1,710 & 1,078 & 4,621 & 3,614\\
6 & C & 189 & 73.9 & 789 & 118 & 1,698 & 279 & 886 & 104 & 1,729 & 1,289 & 10,156 & 3,385\\
7 & N & 192 & 59.3 & 895 & 114 & 1,799 & 261 & 695 & 112 & 1,327 & 1,493 & 4,763 & 3,573\\
8 & C & 193 & 49.7 & 578 & 166 & 1,504 & 246 & 694 & 109 & 1,608 & 1,518 & 12,491 & 3,510\\
9 & N & 180 & 66.6 & 716 & 172 & 1,413 & 329 & 693 & 165 & 1,901 & 1,602 & 11,010 & 3,558\\
10 & C & 191 & 57.4 & 982 & 155 & 2,186 & 376 & 1,003 & 150 & 1,806 & 1,450 & 4,308 & 3,632\\
\bottomrule
\end{tabular}
\end{table*}

\end{document}